\documentclass{aastex62}
\usepackage{tabularx}
\received{September 12, 2018}
\revised{October 12, 2018}
\accepted{\today}
\submitjournal{AJ}

\shorttitle{polarization of WASP-18 system}
\shortauthors{Bott et al.}

\begin{document}

\title{The polarization of the planet-hosting WASP-18 system}

\correspondingauthor{Kimberly Bott}
\email{kimbott@uw.edu}

\author[0000-0002-0786-7307]{Kimberly Bott}
\affil{University of Washington Astronomy Department\\
Box 351580, U.W.\\
Seattle, WA 98195 USA\\}
\affiliation{Virtual Planetary Laboratory\\
Box 351580, U.W.\\
Seattle, WA 98195 USA}
\affiliation{University of Washington Astrobiology Program\\
Box 351580, U.W.\\
Seattle, WA 98195 USA}

\author{Jeremy Bailey}
\affiliation{School of Physics, UNSW Sydney\\
Kensington, Sydney, NSW 2052, Australia}
\affiliation{Australian Centre for Astrobiology, UNSW Sydney\\ 
Kensington, Sydney, NSW 2052, Australia}

\author{Daniel V. Cotton}
\affiliation{School of Physics, UNSW Sydney\\
Kensington, Sydney, NSW 2052, Australia}
\affiliation{Australian Centre for Astrobiology, UNSW Sydney\\ 
Kensington, Sydney, NSW 2052, Australia}

\author{Lucyna Kedziora-Chudczer}
\affiliation{School of Physics, UNSW Sydney\\
Kensington, Sydney, NSW 2052, Australia}
\affiliation{Australian Centre for Astrobiology, UNSW Sydney\\ 
Kensington, Sydney, NSW 2052, Australia}

\author{Jonathan P. Marshall}
\affiliation{Academia Sinica, Institute of Astronomy and Astrophysics,\\ 11F Astronomy-Mathematics Building, NTU/AS campus,\\ No. 1, Section 4, Roosevelt Rd., Taipei 10617, Taiwan}

\author{Victoria S. Meadows}
\affil{University of Washington Astronomy Department\\
Box 351580, U.W.\\
Seattle, WA 98195 USA\\}
\affiliation{Virtual Planetary Laboratory\\
Box 351580, U.W.\\
Seattle, WA 98195 USA}
\affiliation{University of Washington Astrobiology Program\\
Box 351580, U.W.\\
Seattle, WA 98195 USA}



\begin{abstract}

We report observations of the linear polarization of the WASP-18 system, which harbors a very massive ($\sim$ 10
M$_J$)  planet orbiting very close to its star with an orbital period of 0.94 days. We find the WASP-18 system is polarized at
$\sim$200 parts-per-million (ppm), likely from the interstellar medium predominantly, with no strong evidence for phase dependent modulation from reflected light from the planet. We set an upper limit of 40 ppm (99$\%$ confidence level) on the amplitude of a reflected polarized light planetary signal.  We compare the results with models
for a number of processes that may produce polarized light in a planetary system to determine if we can rule out any phenomena with this limit.  Models of reflected light from thick clouds can approach or exceed this limit, but such clouds are unlikely at the high temperature of the WASP-18b atmosphere.  
Additionally, we model the expected polarization resulting from the transit of the planet across the star and find
this has an amplitude of $\sim$1.6 ppm, which is well below our detection limits. We also model the polarization due to the tidal distortion of  the star by the massive planet and find this is also too small to be measured currently. 

\end{abstract}



\section{Introduction} \label{sec:intro}

Hot Jupiters are often thought to provide ideal cases to study polarization from exoplanet atmospheres.  Since the amount of light being scattered off a planet's atmosphere will increase with radius (scattering disk) and inversely with the distance to the star (incident flux), for a given atmosphere a larger hot Jupiter orbiting more closely to its star should produce a better signal.  Furthermore a hotter star will produce greater flux in blue wavelengths which are preferentially Rayleigh scattered, a polarizing effect.  
\citet{Seager2000} modelled the expected polarization levels for hot Jupiter-type systems and predicted that linear polarization varying over the orbital cycle at the tens of parts-per-million level might be present in the combined light of the star and planet.  The star---particularly if an inactive FGK dwarf---is typically expected to not contribute much, if any of this polarized light \citep{kemp87,stam04, cotton17a}.
Rayleigh scattering from small cloud particles is anticipated to be the dominant source of polarization for a hot Jupiter contributing to this signal \citep{Seager2000, bailey18} and hence the observation of polarization would be a strong indicator of the presence of clouds; the presence of clouds increasing the albedo may also be important \citep{fauchez17}.  Detection of polarized light from a hot Jupiter can confirm or provide the color of the planet \citep{berdyugina11}, the nature of its bulk atmospheric composition (e.g. strong or weak Rayleigh scatterers, high clouds, no clouds or haze \citep{Seager2000, stam04, fauchez17}), and the orbital parameters \citep{fluri10} even without imaging the planet.

Despite the advantages of using polarimetry to detect and characterize planets, no polarization signal from a hot Jupiter has been detected to date.
Past attempts at polarized light detection such as those described in \citet{lucas09} have not detected significant planetary polarization signals and the initial report of polarized light from HD 189733b has not been independently confirmed \citep[see:][]{berdyugina11, wiktorowicz15b, Bott2016}.
Modern polarimeters currently available such as HIPPI \citep{bailey15} and POLISH2 \citep{wiktorowicz15a} are now capable of detecting polarization at the parts-per-million level, which should be sufficent for detecting Rayleigh scattering polarization signals from hot Jupiters \citep{Seager2000, hough03}.
If the predicted polarized light signals from hot Jupiter exoplanets continue to be undetectable this would suggest either a fundamental misunderstanding of their atmospheres, or a problem with the methodology in observing them in polarized light.  

An important step in making polarimetry a viable approach to exoplanet characterization is the consideration of the many
environmental factors that can contribute to a polarized light signal (e.g. interstellar polarization, circumstellar matter, tidal distortion, stellar activity, transit polarimetry, etc.).  The estimates of polarized light signals from \citet{Seager2000} are based on the Rayleigh scattering of an idealized atmosphere, but other processes from the atmosphere and environment can polarize light while some (like multiple scattering) can dampen the polarization. In the meantime, providing upper limits to a planet's polarized light signal can rule out certain types of planetary phenomena, including clouds \citep{Seager2000, stam04, fauchez17}, if these secondary processes are well characterized.    


Here we report observations of polarization from the WASP-18 (HD 11069) system, which harbors a very massive (10 M$_J$ \citep{Southworth2009}) planet orbiting very close to its star \citep[P $\approx$ 0.94 days,][]{Hellier2009}.

The WASP-18 system may provide an even more promising candidate for polarimetry than the well-studied hot Jupiter HD 189733b which produced a non-detection in polarized light  \citep{wiktorowicz15b,Bott2016} although it was seen as a promising candidate.  Owing to hydrostatic equilibrium, WASP-18b's radius is approximately the same \citep[R $\approx 1.165$ R$_J$,][]{Southworth2009} as less massive hot Jupiters, including HD 189733b \citep[R $\approx 1.138$ R$_J$,][]{Torres2008} .  However the proximity of the planet to the star 
means the planet receives much more light from the star than those hot Jupiters in longer period systems of the same stellar type.  The host star is a late F type, producing more blue light than HD 189733  (an early K dwarf).  In our analysis of the signal we take into consideration the effects of clouds and environmental phenomena which may reduce or complicate the signal. 

\section{Observations} \label{sec:obs}

The WASP-18 system was observed over four observing runs on the 3.9-m Anglo-Australian Telescope (AAT) at Siding
Spring Observatory in Australia. The observations were made with the HIgh Precision
Polarimetric Instrument \citep[HIPPI,][]{bailey15} in ``clear'' mode (i.e. no filter) to allow a maximum signal.  

The dates of observations were August 28--September 2, 2014; June 27th, 2015; October 14th--October 20th, 2015; and
November 30th--December 7th, 2016. Most of these observations have total integration times of 2560--3840 seconds. However,
the October 2015 observations include several shorter integrations of 640 seconds during and around the planet's transit. 
  HIPPI is an aperture polarimeter using a ferroelectric liquid crystal (FLC)
modulator, a Wollaston prism analyzer and two photomultiplier tubes (PMT) as detectors. The FLC provides a 500 Hz
primary modulation which is used together with two additional stages of slower modulation obtained by rotation of the
Wollaston prism and detectors, and finally by rotating the whole instrument to four position angles (0, 45, 90, 135
degrees) using the AAT's Cassegrain rotator. As with our measurements of HD 189733b these redundant angles were
measured to account for instrumental polarization. The sky background signal is subtracted from the data using a 
sky observation made immediately after each science observation at each Cassegrain rotator position.

HIPPI achieves a precision of 4.3 $\times$ 10$^{-6}$ (4.3 parts-per-million or ppm) or better on bright stars
\citep{bailey15}. This is comparable to or better than the precisions reported from polarimeters based on photoelastic
modulators such as the Pine Mountain Observatory polarimeter \citep{kemp81}, PlanetPOL \citep{hough06}, POLISH
\citep{wiktorowicz08} and POLISH2 \citep{wiktorowicz15a,wiktorowicz15b}.  It is sensitive enough to detect polarized
light from a Rayleigh scattering hot Jupiter atmosphere, expected to be on the order of parts- to
tens-of-parts-per-million \citep{Seager2000}.

The detectors were Hamamatsu H10720-210 PMT modules that have ultra-bialkali photocathodes with a peak quantum
efficiency of 43 per cent at 400 nm. The WASP-18 observations were made without a filter in an attempt  to maximize the
signal on this relatively faint object (B = 9.7). In this mode the wavelength response extends from 350 to 700 nm but
is peaked towards the blue end of this range. Using our bandpass model \citep{bailey15} we find an effective 
wavelength of 474 nm for the F6 spectral type of WASP-18.



The telescope introduces a small polarization (telescope polarization or TP) that must be corrected for. As described
in \citet{bailey15} we determine the TP using observations of a number of stars we believe to have very low polarization
either based on previous PlanetPol observations \citep{hough06,bailey10} or because of their small distances and
expected low levels of interstellar polarization \citep[see ][]{bailey10,cotton16} listed in Table \ref{tab_tp}. The telescope polarization has
been found to be stable during each run, but changes each time the telescope mirror is re-aluminized, which is done
every year for the AAT.  Analysis of our WASP-18 data averaged over each run shows no evidence for any changes between run that might be attributed to the TP calibration, and the standard deviation is the same for Stokes Q and U.

\begin{table}
\caption{Low polarization star measurements to determine telescope polarization (TP) in the ``Clear'' setting (no
filter) for the four observing runs referred to in this paper. Theta, $\theta$, refers to the position angle.}
\begin{center}
\begin{tabular}{llrrrr}
\hline
Star &  Date &  \hspace{15 mm} Q/I  & \hspace{15 mm} U/I  &  \hspace{2 mm} $p$ (ppm) &   \hspace{2 mm}$\theta$ ($\degr$) \\
\hline
\hline

HD 2151  &    29 Aug &   \hspace{1.0 mm}30.0 $\pm$ 3.1 & \hspace{-1.5 mm}-41.8 $\pm$ 3.1 & &   \\
HD 2151  &    31 Aug &   \hspace{1.0 mm}34.2 $\pm$ 3.1 & \hspace{-1.5 mm}-47.8 $\pm$ 2.9  & &  \\
\hline
Adopted TP & Aug 2014 & \hspace{1.0 mm}32.1 $\pm$ 2.2 & \hspace{-1.5 mm}-44.8 $\pm$ 2.1  & 55.1 & 117.2 \\
\hline
\hline

HD 48915    &    23 May &   -35.0 $\pm$ 0.8 & -3.2 $\pm$ 0.8  & &  \\
HD 140573  &    22 May &   -39.2 $\pm$ 3.5 &   \hspace{1.0 mm}3.1 $\pm$ 3.4  & &  \\
HD 140573  &    25 May &   -47.5 $\pm$ 9.7 & \hspace{-1.5 mm}-17.4 $\pm$ 11.0  & &  \\
HD 140573  &    27 Jun  &   -38.6 $\pm$ 2.9 & \hspace{1.0 mm}3.7 $\pm$ 3.0  & &  \\
\hline
Adopted TP & May 2015 & -40.1 $\pm$ 1.6 & -3.5 $\pm$ 1.8  & 40.2 & 92.5  \\
\hline
\hline

HD 2151 &   14 Oct  &   -61.9 $\pm$ 3.4 & -4.9 $\pm$ 3.5  & &  \\
HD 2151 &   15 Oct  &   -54.9 $\pm$ 3.2 & -3.0 $\pm$ 3.2  & &  \\
HD 2151 &   16 Oct  &   -56.0 $\pm$ 3.2 & -6.9 $\pm$ 3.1  & &  \\
HD 2151 &   17 Oct  &   -47.7 $\pm$ 3.9 & \hspace{1.0 mm}3.5 $\pm$ 3.9  & &  \\
HD 2151 &   19 Oct  &   -45.4 $\pm$ 3.3 & \hspace{1.0 mm}2.8 $\pm$ 3.2  & &  \\
HD 48915 &   14 Oct  &   -44.2 $\pm$ 1.1 & -3.7 $\pm$ 0.9  & &  \\
HD 48915 &   15 Oct  &   -43.8 $\pm$ 0.7 & \hspace{1.0 mm}3.1 $\pm$ 0.7  & &  \\
HD 48915 &   15 Oct  &   -51.0 $\pm$ 0.8 & -1.1 $\pm$ 0.7  & &  \\
HD 48915 &   16 Oct  &   -52.4 $\pm$ 0.6 & -4.1 $\pm$ 0.6  & &  \\
HD 48915 &   17 Oct  &   -48.6 $\pm$ 8.7 & \hspace{1.0 mm}4.0 $\pm$ 11.3  & &  \\
HD 48915 &   20 Oct  &   -46.9 $\pm$ 4.6 & \hspace{1.0 mm}1.4 $\pm$ 3.2  & &  \\

\hline 
Adopted TP & Oct 2015 & -50.1 $\pm$ 0.4 & -0.8 $\pm$ 0.3 & 50.3 & 90.5  \\
\hline
\hline

HD 2151    &    30 Nov &   -33.9 $\pm$ 3.5 & \hspace{1.0 mm}6.9 $\pm$ 3.4 & &   \\
HD 48915  &    30 Nov &   -23.3 $\pm$ 2.5 & \hspace{1.0 mm}8.8 $\pm$ 2.7  & &  \\
HD 48915  &    1 Dec  &   -20.6 $\pm$ 0.8 & \hspace{1.0 mm}6.3 $\pm$ 0.8  & &  \\
HD 48915  &    1 Dec  &   -17.2 $\pm$ 4.9 & \hspace{-1.5 mm}-21.0 $\pm$ 5.1  & &  \\
HD 48915  &    2 Dec  &   -24.0 $\pm$ 1.3 & \hspace{1.0 mm}1.2 $\pm$ 1.6  & &  \\
\hline
Adopted TP & Dec 2016 & -23.8 $\pm$ 0.4 & \hspace{1.5 mm}0.4 $\pm$ 0.5 & 23.8 & 89.5  \\
\hline
\hline

\end{tabular}
\end{center}
\label{tab_tp}
\end{table} 

Full details of the observation, calibration, and data reduction procedures with HIPPI can be found in
\citet{bailey15}.









\section{Results} \label{sec:results}

Table \ref{tab_results} lists the individual polarization observations of WASP-18, corrected for telescope polarization
using the values in Table \ref{tab_tp}.
We list the mid-point time, and the normalized Stokes parameters Q/I and U/I given in parts-per-million (ppm) on the equatorial system. 
The measurements are corrected for the wavelength dependent
modulation efficiency of the instrument. This is calculated using the bandpass model described by
\citet{bailey15}. The value is close to 81.3 per cent for
all of these observations. 

\begin{table}[h!]
\caption{The fully calibrated WASP-18 polarization observations without binning.}
\begin{tabular}{rrr}
\hline UT Date and Time & Q/I (ppm) &   U/I (ppm)   \\   \hline
 2014-08-28 16:02:41                 & -58.6 $\pm$ 26.1  & 206.4 $\pm$ 25.5 \\
 2014-08-28 17:04:03                 & -99.3 $\pm$ 27.2 & 172.9 $\pm$ 26.2 \\
 2014-08-28 18:10:35                 & -74.7 $\pm$ 23.2  & 152.9 $\pm$ 27.7 \\

2014-08-29 16:31:05                 & -104.5 $\pm$ 25.3  & 164.1 $\pm$ 25.4 \\
2014-08-29 17:32:44                  & -36.8  $\pm$ 25.7  & 167.0 $\pm$ 25.9 \\
2014-08-29 18:35:31                  & -82.1 $\pm$ 26.5  & 194.5 $\pm$ 26.4 \\

2014-08-30 16:04:47                  & -87.6 $\pm$ 25.7  & 178.6 $\pm$ 25.4 \\
2014-08-30 17:05:02                  & -71.1 $\pm$ 25.7  & 187.5 $\pm$ 25.0\\
2014-08-30 18:05:58                  & -61.0 $\pm$ 25.9  & 188.5 $\pm$ 25.5 \\

2014-08-31 15:29:42                  & -39.6 $\pm$ 25.9  & 250.7 $\pm$ 26.0 \\
2014-08-31 16:28:50                  & -45.3 $\pm$ 25.4  & 189.6 $\pm$ 26.1 \\
2014-08-31 17:28:44                  & -77.4 $\pm$ 25.6  & 225.9 $\pm$ 25.5 \\

2014-09-02 17:01:17                  &-96.0 $\pm$ 29.4  & 215.5 $\pm$ 28.8 \\
2014-09-02 18:02:16                  & -78.4 $\pm$ 28.3  & 169.3 $\pm$ 27.7 \\
2014-09-02 18:50:29                  & -74.1 $\pm$ 39.9  & 100.8 $\pm$ 41.4 \\

2015-06-27 17:27:43                  & -85.2 $\pm$ 26.5  & 177.2 $\pm$ 26.8 \\
2015-06-27 18:27:26                  & -80.7 $\pm$ 26.3  & 182.7 $\pm$ 26.6 \\

2015-10-14 13:21:30                  & -93.5 $\pm$ 27.0  & 213.2 $\pm$ 27.4 \\
2015-10-14 14:21:53                  & -48.7 $\pm$ 27.3  & 129.7 $\pm$ 27.6 \\
2015-10-14 15:20:17                 & -42.0 $\pm$ 27.8  & 190.1 $\pm$ 27.4 \\
2015-10-14 16:23:14                  & -73.2 $\pm$ 27.7  & 236.0 $\pm$ 27.9 \\

2015-10-15 13:20:53                  & -48.4 $\pm$ 27.0  & 158.1 $\pm$ 26.9 \\
2015-10-15 14:23:17                  & -61.5 $\pm$ 26.9  & 169.4 $\pm$ 27.2 \\
2015-10-15 15:23:31                  & -37.4 $\pm$ 26.9  & 183.5 $\pm$ 26.5\\
2015-10-15 16:46:18                  & -85.4 $\pm$ 26.7  & 190.6 $\pm$ 24.4\\

2015-10-16 13:14:56                  &-14.5 $\pm$ 26.3  & 165.7 $\pm$ 26.2 \\
2015-10-16 14:13:00                  & -87.2 $\pm$ 26.6  & 210.7 $\pm$ 26.5 \\
2015-10-16 15:11:33                  & -91.7 $\pm$ 26.7  & 180.8 $\pm$ 26.6 \\

\hline
\end{tabular}
\begin{tabular}{rrr}
\hline UT Date and Time & Q/I (ppm) &   U/I (ppm)   \\   \hline

2015-10-16 16:16:59                  & -92.1 $\pm$ 22.7  & 207.7 $\pm$ 22.8 \\

2015-10-17 10:14:36                  & -115.8 $\pm$ 30.8  & 172.4 $\pm$ 30.1 \\
2015-10-17 13:14:14                  & -20.7 $\pm$ 31.8  & 138.1 $\pm$ 35.7 \\

2015-10-19 10:35:11                  & -59.2 $\pm$ 24.9  & 202.2 $\pm$ 26.8 \\
2015-10-19 13:51:05                  & -14.2 $\pm$ 51.4  & 274.7 $\pm$ 52.4 \\
2015-10-19 14:14:18                  & -171.3 $\pm$ 54.4 & 289.6 $\pm$ 53.7 \\
2015-10-19 14:39:50                 & -20.3 $\pm$ 54.6  & 153.2 $\pm$ 54.2 \\
2015-10-19 15:03:15                  & -176.8 $\pm$ 52.4  & 209.2 $\pm$ 51.9 \\
2015-10-19 15:26:07                  & -150.2 $\pm$ 52.0  & 219.7 $\pm$ 52.4 \\
2015-10-19 16:08:19                  & -64.0 $\pm$ 24.6  & 193.4 $\pm$ 26.5 \\

2015-10-20 11:19:37                  & -83.6 $\pm$ 38.7  & 172.2 $\pm$ 39.3 \\
2015-10-20 11:51:32                  & -30.9 $\pm$ 55.7  & 214.4 $\pm$ 56.5 \\
2015-10-20 12:15:20                  & -66.0 $\pm$ 55.8  & 137.7 $\pm$ 55.2 \\
2015-10-20 12:43:01                  & -88.7 $\pm$ 55.0  & 107.9 $\pm$ 55.8 \\
2015-10-20 13:12:52                  & -132.3 $\pm$ 54.5  & 117.3 $\pm$ 54.1 \\
2015-10-20 15:06:19                  & -60.5 $\pm$ 27.9  & 131.3 $\pm$ 27.5 \\
2015-10-20 16:11:16                  & -83.0$\pm$ 23.8  & 154.4 $\pm$ 24.0 \\

2016-11-30 13:26:59                  & -13.5 $\pm$ 33.9  & 160.5 $\pm$ 32.0 \\

2016-12-01 11:00:29                  & -40.0 $\pm$ 29.4  & 224.2 $\pm$ 29.5 \\
2016-12-01 11:58:27                  & -101.1 $\pm$ 30.7  & 202.4 $\pm$ 32.6\\
2016-12-01 12:57:55                  & -54.8 $\pm$ 30.3  & 196.6 $\pm$ 31.3 \\
2016-12-01 13:55:28                  & -59.6 $\pm$ 29.5  & 151.2 $\pm$ 29.6 \\
2016-12-01 14:51:31                  & -47.6 $\pm$ 31.2  & 225.6 $\pm$ 31.4 \\

2016-12-03 10:58:08                  & -96.4 $\pm$ 51.3  & 282.5 $\pm$ 49.1 \\

2016-12-04 11:24:03                  & -43.2 $\pm$ 43.5  & 227.6 $\pm$ 45.1 \\
2016-12-04 12:08:31                  & -21.4 $\pm$ 40.6  & 209.9 $\pm$ 42.0 \\

2016-12-07 15:06:22                  & -71.0 $\pm$ 34.5  &  181.6 $\pm$ 33.5 \\
2016-12-07 16:04:50                  & -68.9 $\pm$ 38.9  & 185.5 $\pm$ 42.1 \\

\hline
\end{tabular}
\label{tab_results}
\end{table}


Orbital phase is calculated according to the ephemeris where zero phase corresponds
to mid-transit.  Our adopted epoch is from \citet{Maxted2013}, the period is from \citet{Wilkins2017}.

\begin{equation}
\mathrm{ T} = \mathrm{ HMJD} \, 2455265.5525 + 0.94145287 \, \mathrm{ E}  .
\end{equation} 

Typically we observed WASP-18 a few times per night; the average error of each of these unbinned data points is 33.3 ppm.  This is on par with \textit{nightly mean} errors from other polarimeters on hot Jupiter systems (e.g. \citet[]{wiktorowicz15b}).


Two important observations can be made from the data: first, there is clear evidence of polarization at a level of $\sim200$ ppm, and second that the unbinned data points are in reasonable agreement with each other.  The latter point suggests that our instrument is reliable, and that there is little contribution from noise sources over short time scales (such as star spots).

In fitting simple Rayleigh curves to the data, there is no significant difference between using the
individual transit points and binned points (by phase). The data points within transit are inherently noisier due to the shorter duration of the observations, but their inclusion does not significantly change the best fit
for the Rayleigh curve.  Although the polarization values vary by over 200 ppm, the best fit curve of the
polarization ($P = \sqrt{Q^{2}+U^{2}}$ ) has an amplitude of 16.9$\pm$9.8 ppm. This 200 ppm variation may be entirely from noise imparted by the many processes that contribute to the polarization, although we show that some portion of the variation may be from a modulated signal in Section \ref{sec_stats}.  Even in the best
circumstances hot Jupiters are not expected to produce signals stronger than a few tens of
parts-per-million, thus this fit would be in the realm of a reasonable detection if it were shown to be from the planet in the future as more polarimetric measurements of the system, or observation on a larger telescope, would drive down the noise.

\begin{figure}
	\includegraphics[angle=0,width=\columnwidth,trim={0 0 0 0},clip]{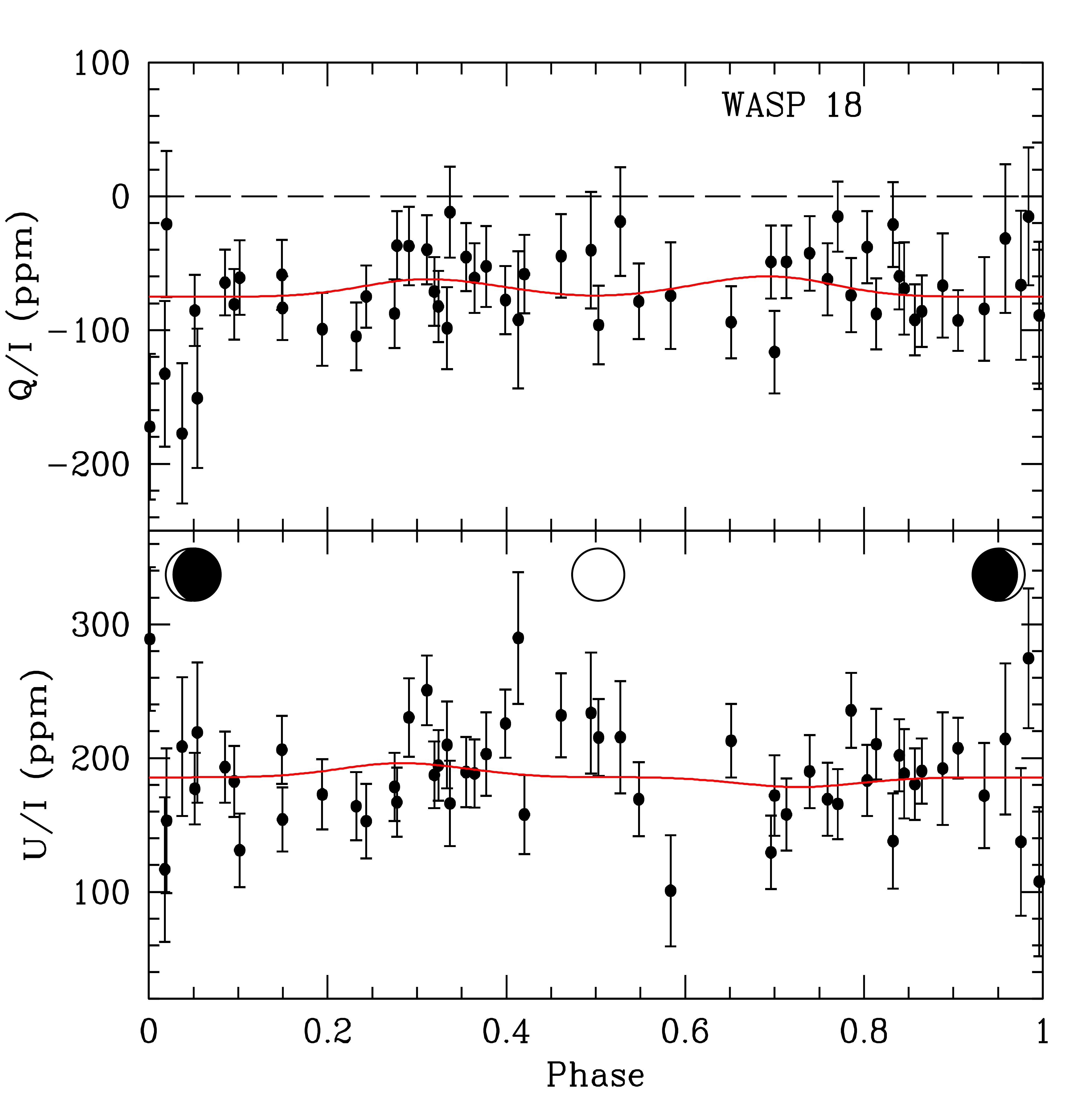}
    \caption{HIPPI measurements of the polarized light from the WASP-18 system including those taken over the transit and secondary eclipse. The data are not binned by run nor by phase. The red curve is a least squares fit of a Rayleigh-Lambert model as described in Section \ref{sec_ray_lamb}. This represents a simplified case where the system polarization would be dominated by Rayleigh scattering in the planet's atmosphere alone.  Note the change in the y-axis for Stokes Q and U.} 
    \label{fig_wasp18}
\end{figure}

Our parameter values for the models throughout this section are taken from \citet{Southworth2009} which derived values for semi-major axis, $a$, planetary radius, R$_\mathrm{P}$, and stellar radius, R$_\star$ comparing observational fits with different stellar models and compares these values to those in literature, finding they are largely in good agreement (with some discrepancy from the \emph{Cambridge} stellar models). The radius of a transiting planet is always contingent upon the certainty of the stellar type and distance to the system, as it is a ratio of the radii (of star and planet) that are retrieved from the normalized reduction in the flux. Our (R$_\mathrm{P}/a)^2$---which is used to scale our direct beam (stellar source) and in the forward radiative transfer and simplified Rayleigh models---is 0.00074.   A gravitational acceleration based on the mass is also used in our forward models for the atmospheric model used as input to the radiative transfer solution; here we also adopt the value from \citet{Southworth2009} of 191 m/s.


\subsection{Rayleigh-Lambert model}
\label{sec_ray_lamb}

Our data is fitted with a Rayleigh-Lambert model for the expected polarization variation. This
is an analytic model which calculates the intensity according to the expected phase variations 
for a Lambert sphere, and assumes the polarization follows the phase function for Rayleigh scattering (see: \citet{Seager2000} and \citet{wiktorowicz09}). To find the best fit we use a
Levenberg-Marquardt non-linear least squares algorithm \citep{press92} with five parameters: the polarization zero point offsets in Q/I and U/I ($Z_q$ and $Z_u$), the polarization
amplitude $p$ which allows the effects of depolarization processes such as multiple scattering to be taken into account, the position angle of the major axis of the projected orbit ellipse on the sky $P\!A$, and the orbital inclination $i$. The fitted parameters and their uncertainties (determined from the covariance matrix
of the fit) are listed in Table \ref{tab_fit}. The fitted model is shown by the red curve in Figure \ref{fig_wasp18}.

\begin{table}
\centering
\caption{Parameters of Rayleigh-Lambert fit to HIPPI linear polarization observations of WASP-18.  $Z_q$ and $Z_u$ are the polarization offsets due to constant sources such as interstellar polarization, $p$ refers to the strength of the polarized light modulation from Rayleigh scattering in the planet atmosphere, $PA$ the position angle offset of the orbital plane, and $i$ its inclination.}
\begin{tabular}{lll}
\hline
Parameter & Value & units \\
\hline
$Z_q$ & -75.2 $\pm$ 6.0& ppm \\
$Z_u$ & 185.0 $\pm$ 5.5 & ppm \\
$p$ & 16.2 $\pm$ 10.0 & ppm \\
$PA$ & 200.3 $\pm$ 20.7 & degrees \\
$i$ & 79.2 $\pm$ 10.9 & degrees \\
\hline
\end{tabular}
\label{tab_fit}
\end{table}

The best fit inclination for this system from our data ($79.2\pm10.9^{\circ}$) is lower than that suggested by the transit and
phase curve model fitting done by \citet{Hellier2009} ($86.0 \pm2.5^{\circ}$), though they overlap, and are hence consistent, within their errors. We can
fit the Rayleigh-Lambert model with the inclination fixed at 86 degrees, and obtain very similar results, with $p = 17.3 \pm
10.2$ ppm. The impact parameter for this system reaches unity (a skimming transit) at an inclination of about 74 degrees. Thus the
fitted model is consistent with the transiting nature of the system.

However, the error on the polarization amplitude is such that the fitted signal cannot be considered a significant detection of
planetary polarization. Furthermore, when we account for statistical bias effects we will show that it is even less significant
than it appears at first sight. 

To improve our sensitivity to such effects we would need either substantially more data, or an instrument like HIPPI on a larger
telescope.  With an 8-meter class telescope we would be able to halve the uncertainties.

\subsection{Statistical Tests}
\label{sec_stats}

A simple way to check for variability in polarimetric data is to calculate the moments of the distribution for the Q and U data and compare the values with tables to determine significance. Table \ref{madeupwords} presents our calculations of standard deviation, skewness and kurtosis for both Q and U; the skewness being the division of the 3\textsuperscript{rd} moment of the data set squared by the second cubed, and the kurtosis being the 4\textsuperscript{th} moment of the data set, divided by the second moment squared. The average uncertainty in our data is 33.3 ppm, a value that represents the internal standard deviation of individual measurements and which scales with photon-shot noise. The standard deviation (or scatter) in both Q and U is higher than 33.3 ppm, which may indicate some real variability. The scale of the additional variability may be calculated as $\sqrt{x^2-e^2}$, where $x$ is the scatter, and $e$ the average error; for this data the values are 10.8 ppm in Q and 20.3 in U, the mean of which is similar to the fitted value of polarization, p in the Rayleigh model. The difference between the scatter and mean uncertainty could be due to a real signal, but it may also be accounted for by differences in centring, minor differences in telescope polarization (TP) calibration between runs or other systematics.  As mentioned in section \ref{sec:obs} there do not appear to be significant TP calibration difference between runs.

Using the significance tables given by \citet{brooks94} for n = 56, we see that in U both skewness and kurtosis are consistent with a Gaussian distribution. The fitted Rayleigh model amplitude is greater in Q than U though, and in Q we see some evidence for non-Gaussian behaviour. In Q the skewness is non-Gaussian at the 95\% confidence level ($0.6097\pm0.0032$), but not the 99\% confidence level (0.8404$\pm$0.0054). The kurtosis doesn't quite meet the threshold for being non-Gaussian at 95\% confidence ($4.2931 \pm 0.0177$), but it is high. In Q the skewness is positive, and the kurtosis high; together this implies a greater number of more positive outlying data points than expected from a normal distribution, and this is consistent with the form of the Rayleigh fit to the Q data shown in Figure \ref{fig_wasp18}.

\begin{table}
\centering
\caption{Moment calculations: the terms are described in the text.}
\begin{tabular}{llll}
\hline
data & std dev & skewness & kurtosis \\
\hline
q & 35.0 & 0.6320 & 4.1286 \\
u & 39.0 & 0.0847 & 3.4414 \\
\hline
\end{tabular}
\label{madeupwords}
\end{table}

When fitting a polarization amplitude to the observations as in the previous section it should be noted that the amplitude is
subject to a statistical bias since the fitted amplitude can never be negative. The effect is analogous to the bias well
known to be present in measurements of the degree of polarization \citep[e.g.][]{simmons85}. The bias is most significant when
the signal-to-noise ratio is small.

\begin{figure}
	\includegraphics[angle=0,width=\columnwidth,trim={0 0cm 0 0cm},clip]{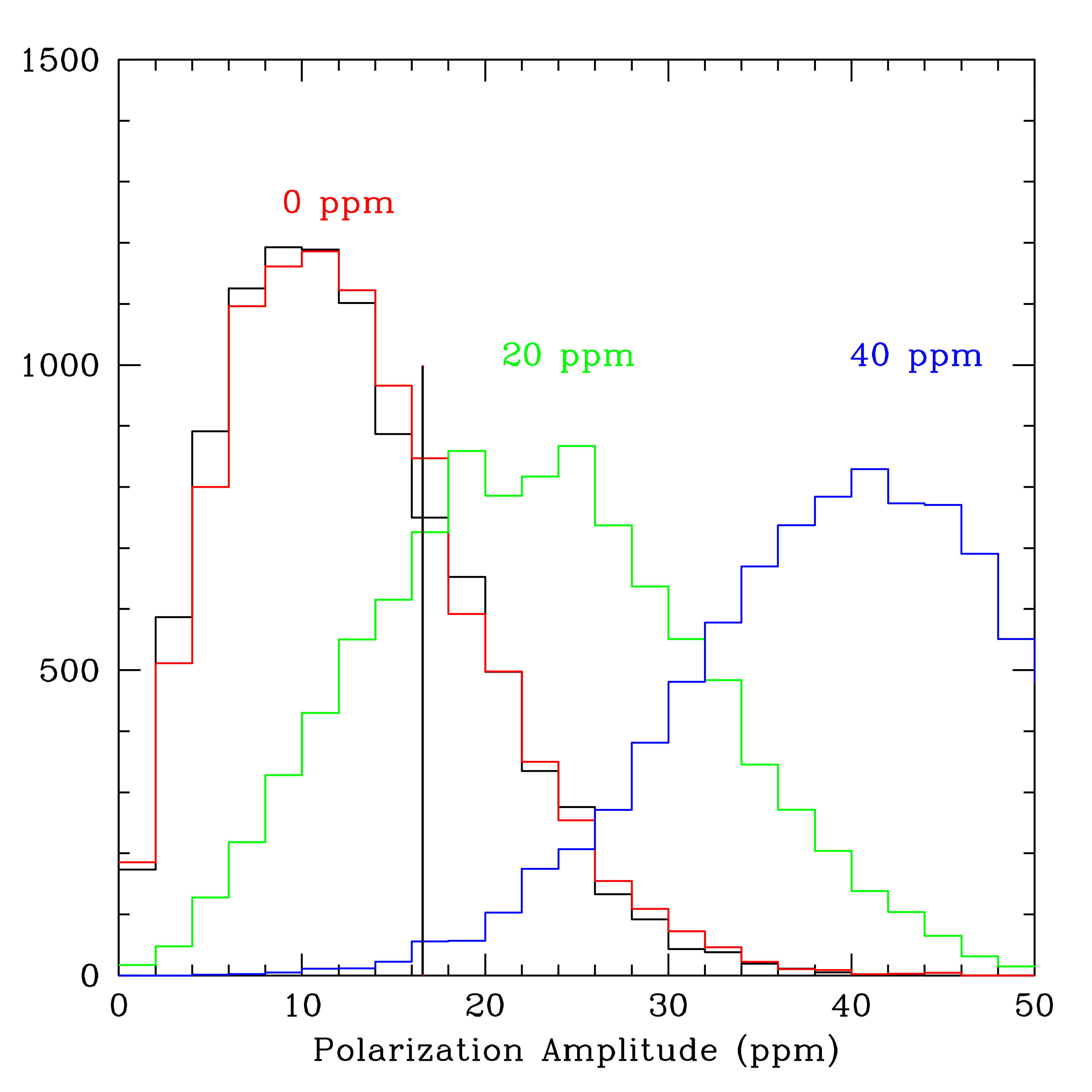}
    \caption{A confidence test with random sampling for detections at the 0 (red), 20 (green) and 40 (blue) ppm level.  The black histogram corresponds to a bootstrap random sampling for 0 ppm.  The 0 ppm samples are offset from a zero ppm amplitude because scattered data points may be fit to a zero amplitude curve when they do not coincide with a Rayleigh phase curve shape.  Our observed polarization amplitude is shown by the vertical black line.  This lies well into the wings of the 40 ppm curve where we can rule out a detection at the 40 ppm level with 99\% confidence.   } 
    \label{fig_liklihood}
\end{figure}

To test the size of this effect we carried out a set of trials where we generated random data sets with the same phasing and
noise properties as our data, and fitted Rayleigh-Lambert models in the same way as just described. The results are shown in
Figure \ref{fig_liklihood}. The black histogram represents a bootstrap sampling where we used the actual data points but randomly swapped
them in phase. The red histogram represents data generated with a zero amplitude signal but noise randomly added in line with the
errors of the actual data points. Both of these tests which were run for 10,000 trials generated very similar histograms, with a
mean fitted amplitude of about 12 ppm. Of these trials 27\% produced fitted amplitudes greater than the 16.2 ppm amplitude
fitted to the actual data even though there was zero true signal in the generated data.

The green and blue histograms are for cases where the injected signal had amplitudes of 20 ppm and 40 ppm respectively.  The 40
ppm injected signal produced fitted amplitudes higher than our 16.2 ppm signal, more than 99\% of the time. Hence we can rule
out a planetary signal of 40 ppm or greater at the 99\% confidence level.   

\subsection{Constant polarization}

A few different processes can lead to a polarized light signal outside of the modulated signal from the planet.  In the cases of contributions from zodiacal dust and the interstellar medium (ISM) they are separable from the planetary signal in theory because they produce a constant signal, an offset in Stokes Q and U from zero, however these still contribute to the noise in the signal.

In our paper on HD 189733 \citep{Bott2016}, an offset in the modulated signal from zero was detected.  This background
polarization can be due in large part to interstellar polarization from the interstellar medium \citep{bailey10,
cotton16} and hence can be approximated if the polarization of the ISM is well characterized.  WASP-18 is 126.4 parsecs from Sol \citep{gaia16}, at galactic coordinates 279.7100, -69.3214.  Considering the trends in interstellar polarization from \citet{cotton16}, we can estimate the expected offset from modulation around
zero for the polarized light to be approximately 200 ppm for the system. This is rather more than the values we found for HD 189733, but appropriate to the measurement of this system (see Table \ref{tab_fit}). 

At the distance of WASP-18 the model described in \citet{cotton16} is informed by very little data. This distance is near the wall of the local hot bubble, a region of space with a greater dust density, imparting greater interstellar polarization \citep{leroy99}. So, as a further test of the level of interstellar polarization we observed three stars near to WASP-18 in the same bandpass. These stars were chosen for: having spectral types not known to be intrinsically polarized, their brightness, and their proximity to WASP-18. Table \ref{polcontrolstars} gives the results of these observations. The differences between the control stars, likely due to the patchiness of the interstellar medium,  prevents us from precisely determining the interstellar polarization for WASP-18 directly from these measurements. However, it is clear the control stars have a similar magnitude and polarization position angle to that which may be calculated from $Z_q$ and $Z_u$ in Table \ref{tab_fit} ($Z_p = 199.7\pm5.8$ ppm, $Z_{PA} = 56.1\pm0.9$ deg). Thus it is very likely that the majority of the constant polarization signal is due to interstellar polarization.

\begin{table}
\centering
\caption{Observations of control stars for WASP-18. Spectral type, distance, and position have been taken from SIMBAD. ``Sep'' is the angular separation from WASP-18.}
\begin{tabular}{rrrrrrrrrr}
\hline
Object & SpT & dist(pc) & Sep(deg) & Observed (UT) & Exp(s) & Q (ppm) & U (ppm) & p (ppm) & PA (deg) \\
\hline
HD 9414 & A1V & 99.6 & 1.0134 & 2015-10-16 17:53:47 & 640 & -41.0$\pm$13.2 & 186.7$\pm$13.3 & 191.2$\pm$13.3 & 51.2$\pm$2.0\\

HD 9733 & G8IV & 122.0 & 0.5387 & 2015-10-17 11:43:06 & 1280 & 40.7$\pm$15.0 & 172.6$\pm$14.8 & 177.3$\pm$14.9 & 38.4$\pm$2.4\\

HD 10162 & F0IV & 100.2 & 3.3561 & 2015-10-17 14:50:26 & 1280 & -93.8$\pm$23.6 & 107.8$\pm$19.7 & 142.9$\pm$21.9 & 65.5$\pm$4.4\\
\hline
\end{tabular}
\label{polcontrolstars}
\end{table}

\begin{figure}
	\includegraphics[angle=0,width=\columnwidth,trim={0 7cm 0 7cm},clip]{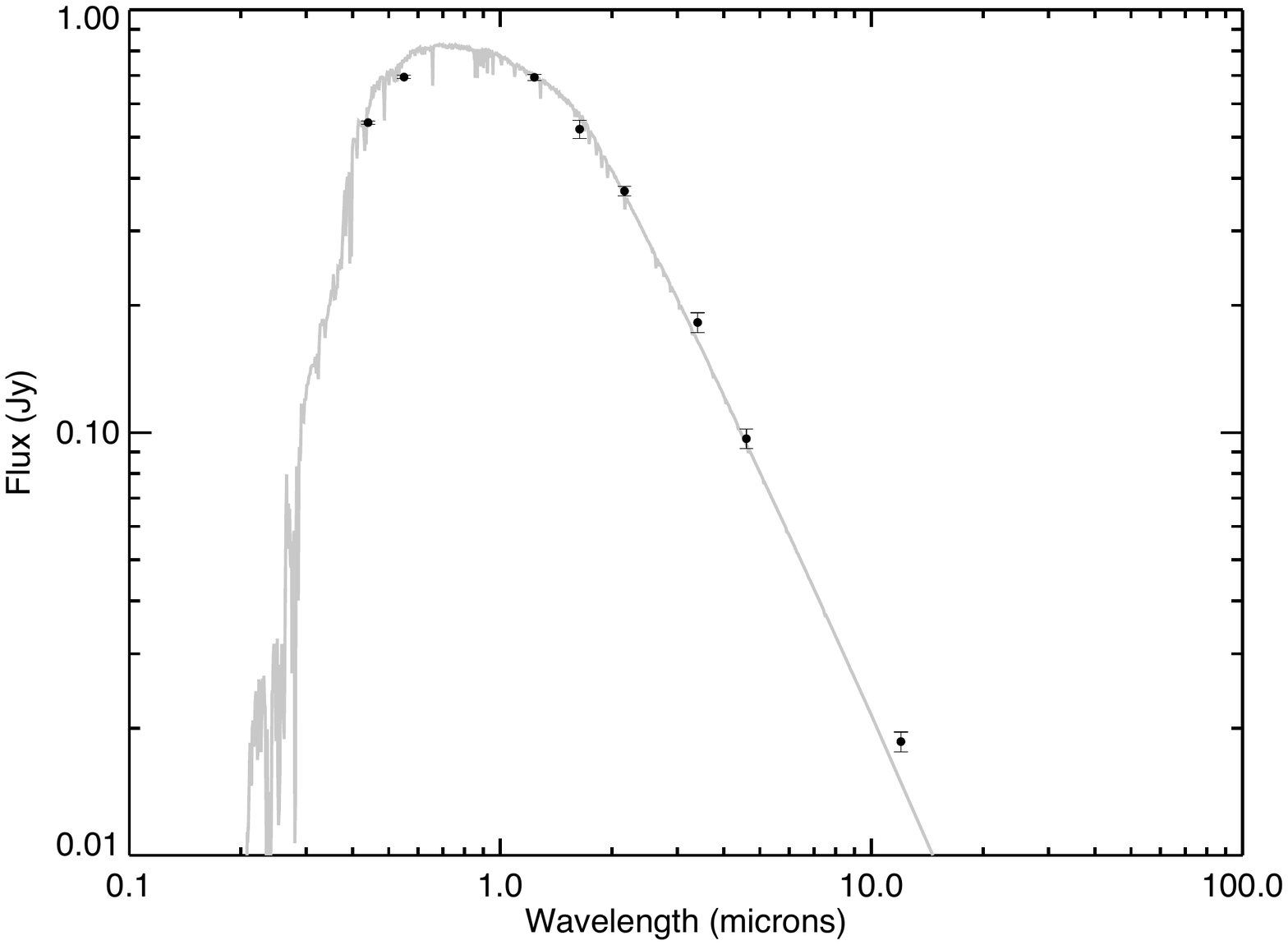}
    \caption{An SED (spectral energy distribution) fit to combined photometry of WASP-18.  Black data points are literature values compiled from optical, near-infrared and mid-infrared catalogues, with 1-sigma uncertainties. The grey line denotes the stellar photosphere model. Note that the photosphere model has not been scaled to fit the photometry - see text for details. No significant excess emission is detected at any wavelength.} 
    \label{fig_sed}
\end{figure}

Circumstellar dust
can also produce a constant polarization signal \citep{kolokolova10, cotton16, marshall16} and while its signal is typically assumed to be constant, the characterization and consideration of the influence of debris is vital to characterizing planets both in polarized and unpolarized observations \citep{roberge12}.  WASP-18 likely does not have any significant
circumstellar debris, although even zodiacal dust can contribute a small degree of polarization offset. In Figure \ref{fig_sed} we show the spectral energy distribution (SED) fit to photometry including the Optical BV (Hipparcos catalogue, \citep{perryman97}), and near-infrared 2MASS JHKs \citep{skrutskie06}, and WISE mid-infrared \citep{wright10}. The same model stellar photosphere from the \textsc{atlas9} database used in our polarization models of transit (Section \ref{transit_pol}) and tides (Section \ref{sec:tidaldist}) is placed on the image with these data points.  The spectral energy distribution (SED) shows no significant excess across the available wavelength range although constraints at and beyond the mid-infrared ($\geq 20\mu$m) are weak.

Furthermore we might consider that such an irradiated planet might be experiencing atmospheric loss which could contribute to the polarization signal. 
While other hot Jupiters have shown evidence of these processes (e.g. HD 189733b \citep{Bourrier2013},
HD 209458b \citep{Linsky2010}, WASP 12b \citep{Haswell2012}) the mass loss estimates of WASP-18's hot
Jupiter are quite small based on the X-ray emission of its host star \citep{Salz2015}.  Contributions to variations in polarization from asymmetries in the planet's atmosphere, such as cyclonic spots and localized hazes, are expected to produce signals of only a few percent \citep{karalidi13}, which is negligible for this planet.  Asymmetries in polarization from the hot spots observed on some hot Jupiters are currently poorly understood.

\subsection{Stellar Activity and Interactions}

A very young star might be expected to have activity leading to polarized light signals from abundant starspots upsetting the symmetry of the star's atmosphere \citep{kostogryz15} or from differential saturation from Zeeman splitting of spectral lines (see \citet{cotton17a}).  The age of the WASP-18 system is poorly constrained, but likely to be fairly young \citep{Hellier2009, Brown2011}.  Isochrone fitting for the star suggests it is 600 Myr, however the non-detection of X-rays from a star that old is unusual, furthermore the slow rotation of the star would imply it is much older than other metrics would suggest.  Interactions with the massive and closely orbiting planet could prematurely slow the star, stealing away rotational energy.  It is also possible that the planet disrupts the stellar magnetic dynamo created within the star's thin convective layers, reducing the mixing and allowing the detected lithium---an indication of youth---to last longer.  Whatever the star's age may be, its current state as indicated from studies of Ca II H and K lines and X-ray activity suggests it is relatively inactive, likely with a weak magnetic field, producing few star spots and not interacting with its planet's magnetic field \citep{miller12}. 

The interactions between the magnetic fields of closely orbiting hot Jupiters and their stars are possible sources of
polarized light \citep{Bott2016, fares10, walker08}.  However, in the WASP-18 system the star is expected to have a weak magnetic field due to tidal interactions with the planet disrupting convection which in turn disrupts the stellar dynamo \citep{pillitteri14}.


\subsection{Transit Polarimetry}
\label{transit_pol}

Our observations cover two transits observed on 2015 October 19th and 20th. 
During the transits we reduced the integration times of our observations to attempt to detect any polarimetric
signals due to the transit. The reduced integration time has the effect of increasing the
uncertainty of these points to around 50 ppm or higher. 

\begin{figure}
	\includegraphics[angle=0,width=\columnwidth,trim={0.7cm 7.0cm 2.0cm 8.5cm},clip]{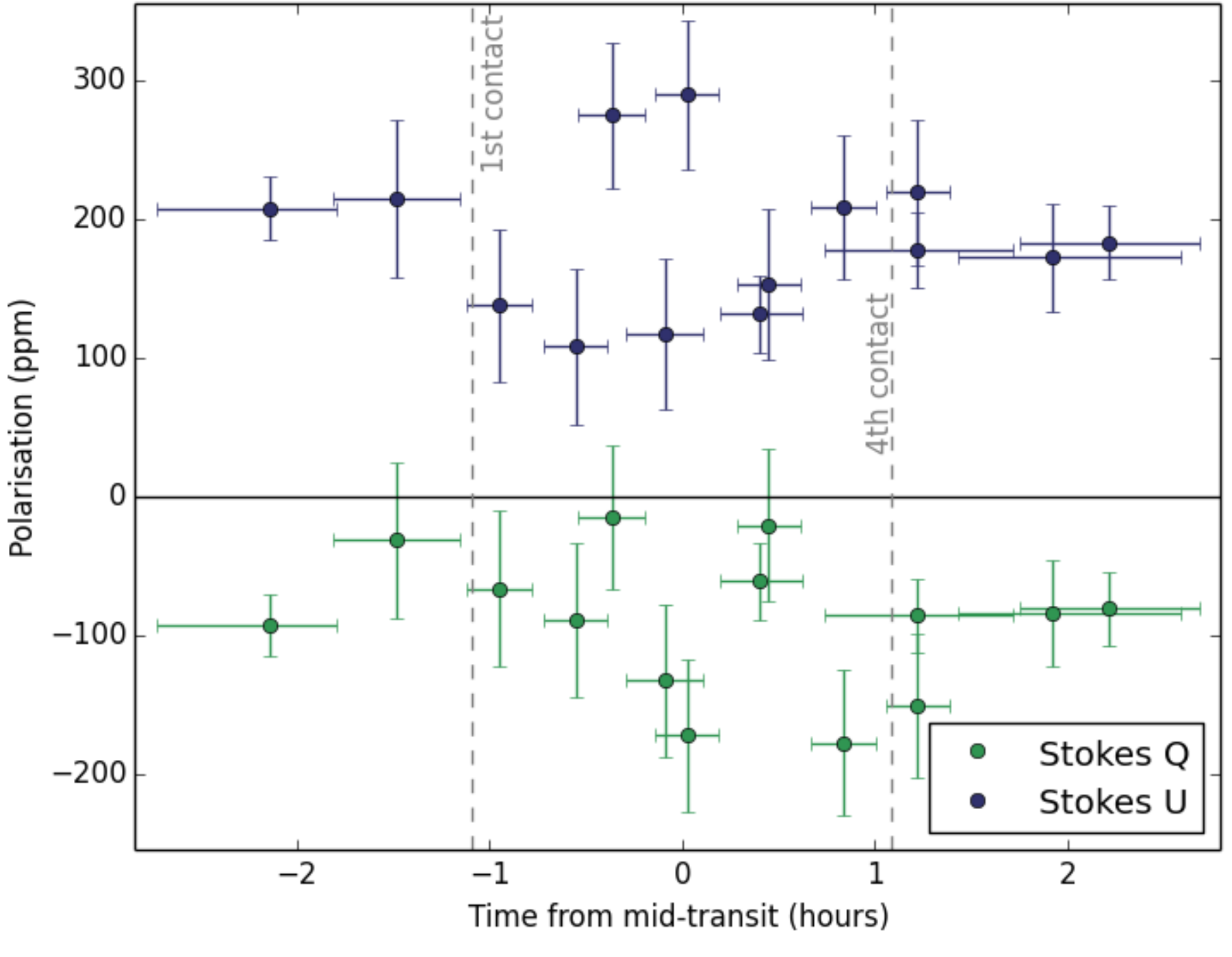}
    \caption{HIPPI measurements of the polarized light from the WASP-18 system around and through transit (1st to 4th contact) centered on mid-transit. Here the data are plotted on the same axes.  The vertical error bars are the true error, while the horizontal error bars show the duration of the integration (start to stop time of observation).  There is more noise in the transit than outside of it because the exposures are shorter.} 
    \label{fig_transit}
\end{figure}

The observations within the transit
are shown in Figure \ref{fig_transit}. These observations show increased scatter compared with the longer,
out-of-transit integrations, but this is consistent with the increased uncertainties. There is no indication of
polarization variability resulting from the transit.

Polarization during transit is expected to be produced as a result of scattering in the
stellar atmosphere. Normally this polarization is perpendicular to the radius vector and increases toward the limb of
the star. The symmetry of a spherical star will cause the polarization to integrate to zero over the whole star.
However, the occultation by a planet will break this symmetry and result in net polarization, which is expected to be
largest when the transiting planet is near the limb of the star. The effect is analogous to the Chandrasekhar Effect
originally predicted for hot stars in eclipsing binaries \citep{chandrasekhar46} and observed in the case of Algol
\citep{kemp83}.  The light transiting through the planet's atmosphere is also affected by polarization \citep{dekok12} but for this geometry of a large gas giant very near its star, the stellar contribution is likely dominant.

This transit polarization effect has been investigated for planets transiting cool stars by \citet{carciofi05} and \citet{kostogryz11}.
\citet{kostogryz15} have calculated the transit polarization for a range of transiting planets, and shown that the size
of the effect increases for lower stellar temperatures, lower gravities, and larger planet to star radius ratios, with
the effect reaching up to 10 ppm or more at 400 nm in the most favourable cases. However, WASP-18 is outside the range of 
parameters considered in that study.

\begin{figure}
    \includegraphics[width=\columnwidth]{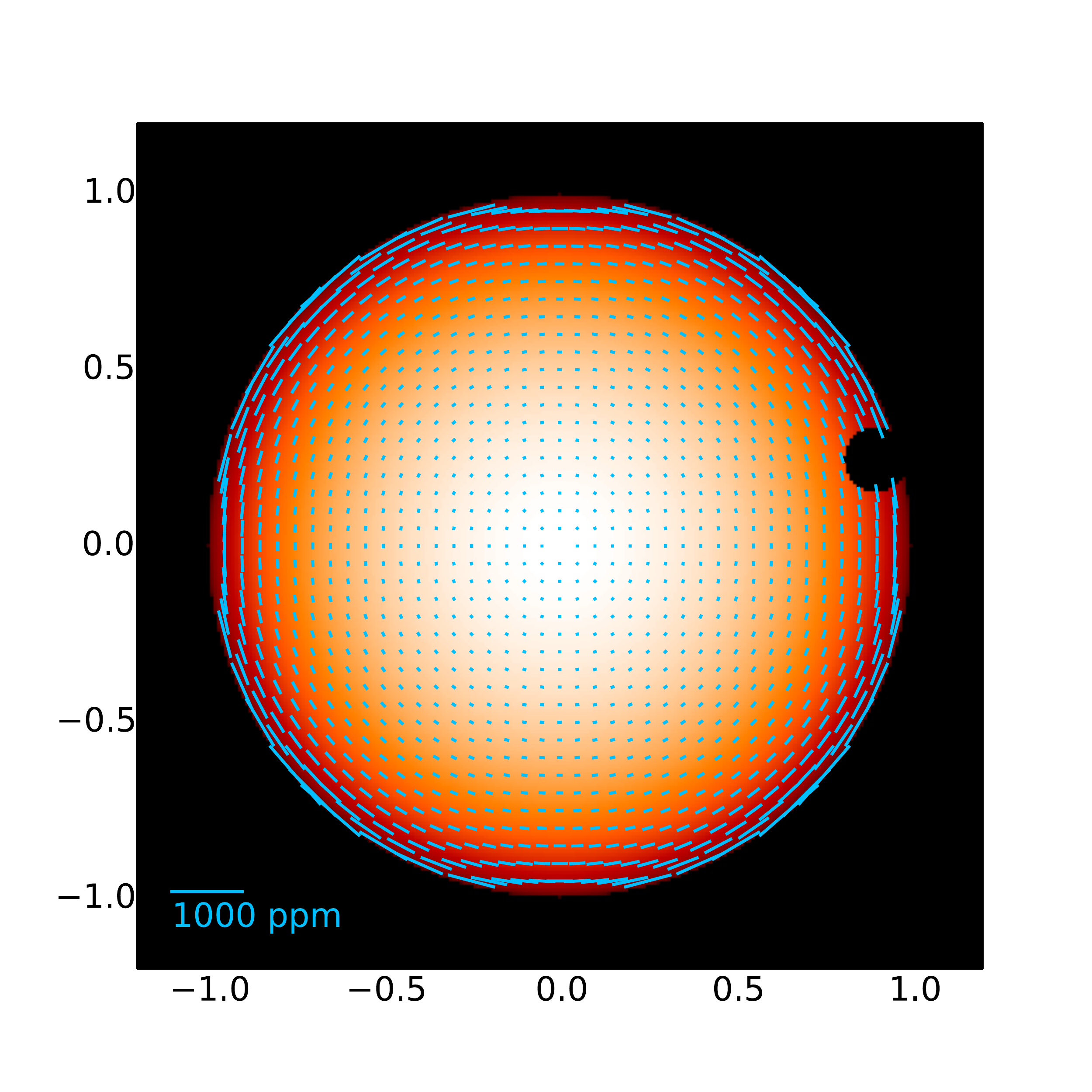}
    \caption{Example of polarization transit modelling. The intensity distribution and overlaid polarization vectors at
    440 nm wavelength are shown for a WASP-18 model with a transiting planet near the phase that produces the maximum
    polarization.}
    \label{transit_pol_ex}
\end{figure}

We have calculated new models for transit polarization over a wider range of stellar types, using an adaptation of the
modelling methods used by \citet{cotton17b} to calculate the polarization due to the rotational oblateness of Regulus.
We start with an  \textsc{atlas9} stellar atmosphere model. For the WASP-18 case we use a model for T$_{eff}$ = 6368 K
and log $g$ = 4.37  \citep[the stellar parameters of WASP-18 according to][]{torres12} based on the solar abundance grid
of  \citet{castelli04}. We then calculate the emergent specific intensity and polarization as a function of wavelength
and viewing angle ($\mu = \cos{\theta}$, where $\theta$ is the local zenith angle) using the \textsc{synspec/vlidort}
code. This is a version of the \textsc{synspec} spectral synthesis code \citep{hubeny85} which we have modified to do
polarized radiative transfer using the Vector Linearized Discrete Ordinate Radiative Transfer (\textsc{vlidort}) code
of \citet{spurr06}. The calculations include the polarization due to Thompson scattering from electrons (most important
in hot stars) and Rayleigh scattering due to H and He atoms and H$_2$ molecules (important in cool stars).

\begin{figure}
    \includegraphics[width=\columnwidth]{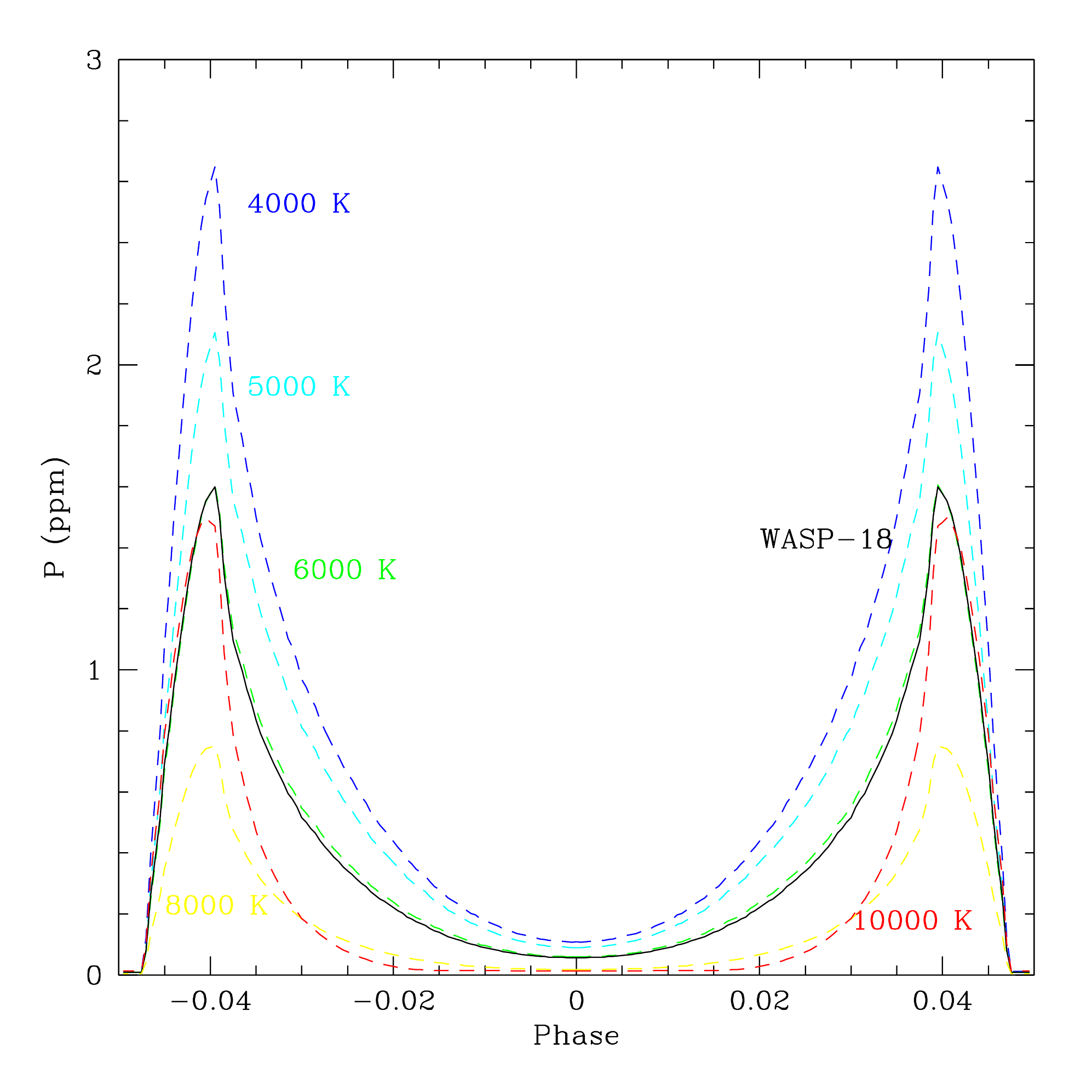}
    \caption{Modelled polarization curves for the WASP-18 transit (solid line) at a wavelength of 440 nm. Dashed lines show models
    for the WASP-18 transit geometry but for stellar atmosphere grid models of different temperatures at log $g$ = 4.5 .}
    \label{tphase_models}
\end{figure}    

We then set up a grid of ``pixels'' covering the observer's view of the star and
spaced at 0.01 of the stellar radius. For each pixel we determine the viewing angle $\mu$ and then interpolate in our
set of \textsc{synspec/vlidort} calculations to determine the intensity and polarization spectrum for that pixel. The
resulting data provide a map of the intensity and polarization distribution across the star at each wavelength. To
model a transit we remove from the map those pixels that are occulted by the planet and sum the remaining pixels to
give the integrated intensity and polarization at that point during the transit \footnote{Since we know the
polarization over the whole star should integrate to zero, it is equivalent to integrate only the pixels
occulted by the planet and then reverse the sign. In practice we use both methods as a consistency check.}. An example of the
resulting data is shown in Figure \ref{transit_pol_ex}.

\begin{figure}
    \includegraphics[width=\columnwidth]{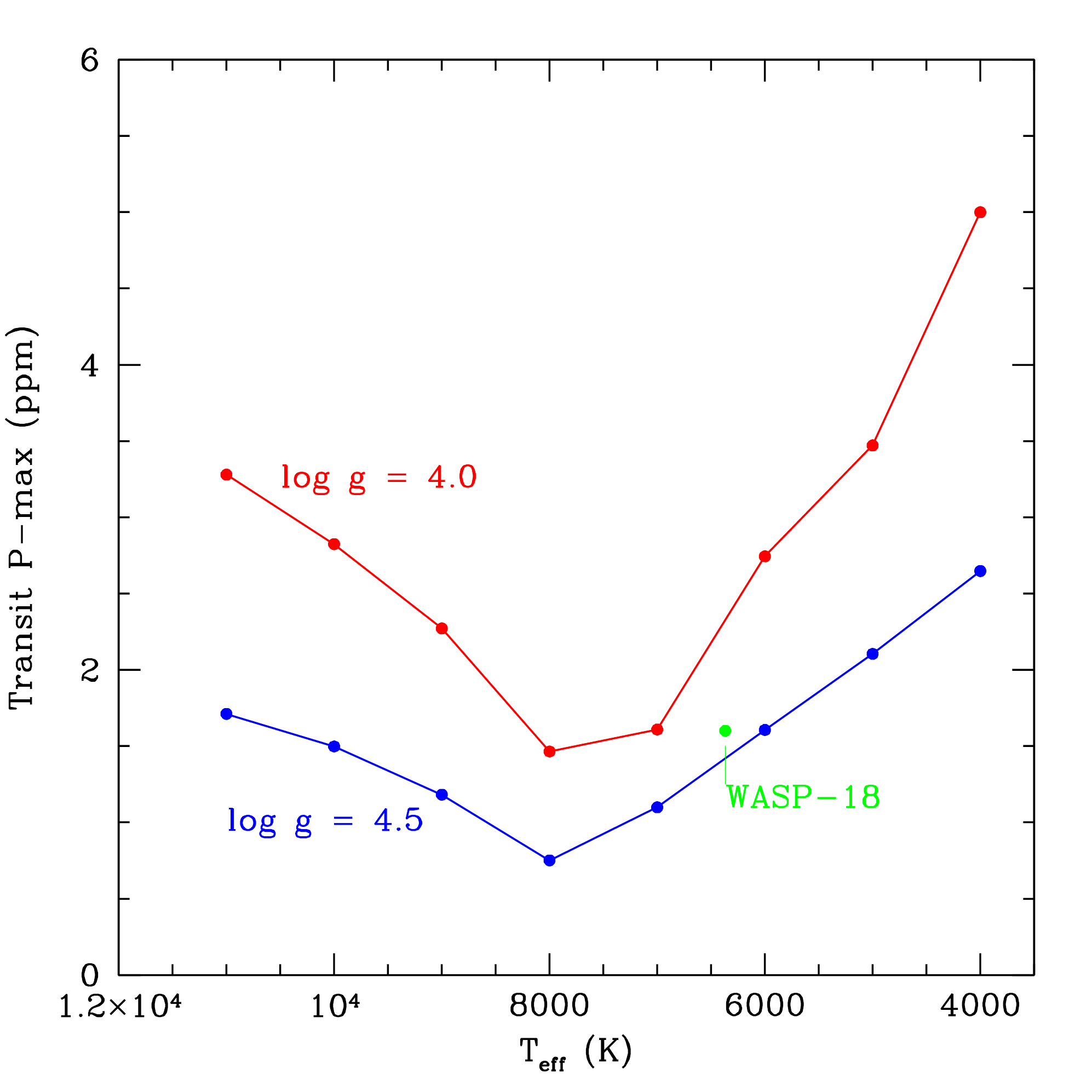}
    \caption{Maximum polarization during transit (at 440 nm) as a function of stellar effective temperature and gravity for the WASP-18 - WASP-18b radius ratio. Grid
    models with log $g$ of 4.0 and 4.5 are shown, while the WASP-18 model with $T_{eff}$ = 6368 K and log $g$ = 4.37 is
    indicated by the green dot. The noise in the transit in Figure \ref{fig_transit} is significantly greater than this.}
    \label{transit_pmax}
\end{figure}

For the results presented here we have modelled the transit geometry appropriate to WASP-18b using an inclination of 
86 degrees (the inclination will change the shape of the curve but not the maximum polarization) and a planet/star radius ratio of 0.0913. Figure \ref{tphase_models} shows results for the polarization variation
during transit for the WASP-18 model shown by the solid line. The polarization amplitude is 1.6 ppm.  Given that our
transit observations have errors of $\sim$50 ppm, and that the scatter in the data is particularly large during the short observations taken during transit (see Figure \ref{transit_pol_ex}), it is unsurprising that we did not detect any effect. 

In addition to the model specific to WASP-18 we have investigated the variation of the transit signal with stellar type
by running models for the WASP-18 transit geometry but substituting stellar atmosphere models taken from the
\citet{castelli04} grid with $T_{eff}$ from 4000 K to 11000 K and log $g$ of 4.5 and 4.0. Some of these models are
shown as dashed lines on Figure \ref{tphase_models}, and in Figure \ref{transit_pmax} we show the maximum polarization
during a transit as a function of temperature and gravity. The results show that the transit polarization signal
reaches a minimum at about 8000 K, but then starts to increase again at higher temperatures, which is due to the
increasing importance of electron scattering in hotter atmospheres.

The recent discovery of transiting planets such as KELT-9b \citep{gaudi17} which orbits a B9.5-A0 star of 10170 K shows
that the transit polarization effect may be observable in hotter stars as well as cool stars.

\subsection{Tidal Distortion}
\label{sec:tidaldist}

Another way of breaking the polarization symmetry of a star is through tidal distortion of the star by the planet and the planet by the star in turn. This
has been suggested as a possible polarization mechanism for exoplanet systems \citep[e.g.][]{hough03} but estimated to be a small effect in
most cases. WASP-18b is, however, an extreme case in view of the mass of the planet, and the closeness of the planet
to the star.

\begin{figure*}
    \includegraphics[width=18cm]{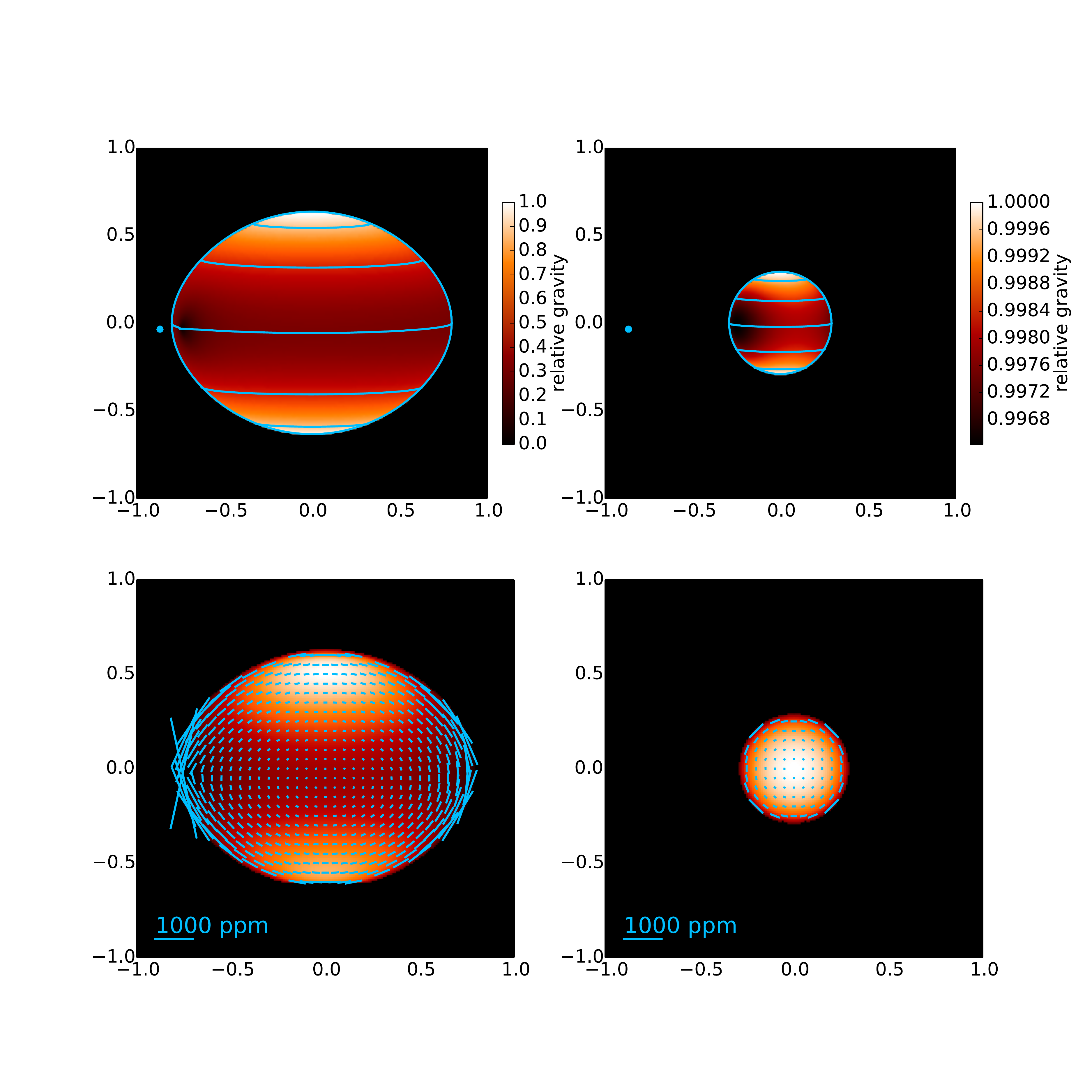}
    \caption{Models for polarization due to tidal distortion. The inclination is 86 degrees and the phase angle is 60 degrees. The left panels show results for a star large enough to fill its Roche lobe, while the right panels are for the actual WASP-18 star size and a rotation rate 5 times slower than the orbital period. At the top is shown the shape and effective gravity distribution over the star. For the Roche lobe filling case there is a low gravity region near the L1 point which results in a cold spot below the planet. This results in an asymmetry that produces a phase dependent polarization effect. For the actual WASP-18 cold spots are seen on either side of the star, but the contrast in gravity is now very small. The lower panels show the resulting intensity distribution and overlaid polarization vectors for a wavelength of 440 nm. The blue dot in the top panels is the planet shown at its correct position. 
}
    \label{tidal_pol}
\end{figure*}

We can model the polarization produced by tidal distortion of the star using the same methods as those in \citet{cotton17b} and in Section \ref{transit_pol} above. We use the formulation of \citet{wilson74} as updated by \citet{wilson79} to describe the geometry of a tidally distorted star including the case of non-synchronous rotation. This allows us to determine the shape of the star, the variation of effective gravity across its surface, and the viewing angle $\mu$ and rotation angle $\xi$ at any point on the star for a given inclination and phase angle. We can then derive intensity and polarization maps and integrated polarizations using the same methods as in \citet{cotton17b}. We use a set of \textsc{atlas9} models covering the required range of gravities and temperatures. We calculate the polarization and intensity for these models using \textsc{synspec/vlidort} and interpolate in between them to obtain results for each ``pixel'' across the star. We assume effective temperature and gravity are related through the von Zeipel gravity darkening law  $T_{eff} \propto g^{\beta}$ where $\beta$ = 0.25 \citep{vonzeipel24}. According to \citet{espinosa12} this is a good approximation for most binary parameters.

\begin{figure}
     \includegraphics[width=\columnwidth]{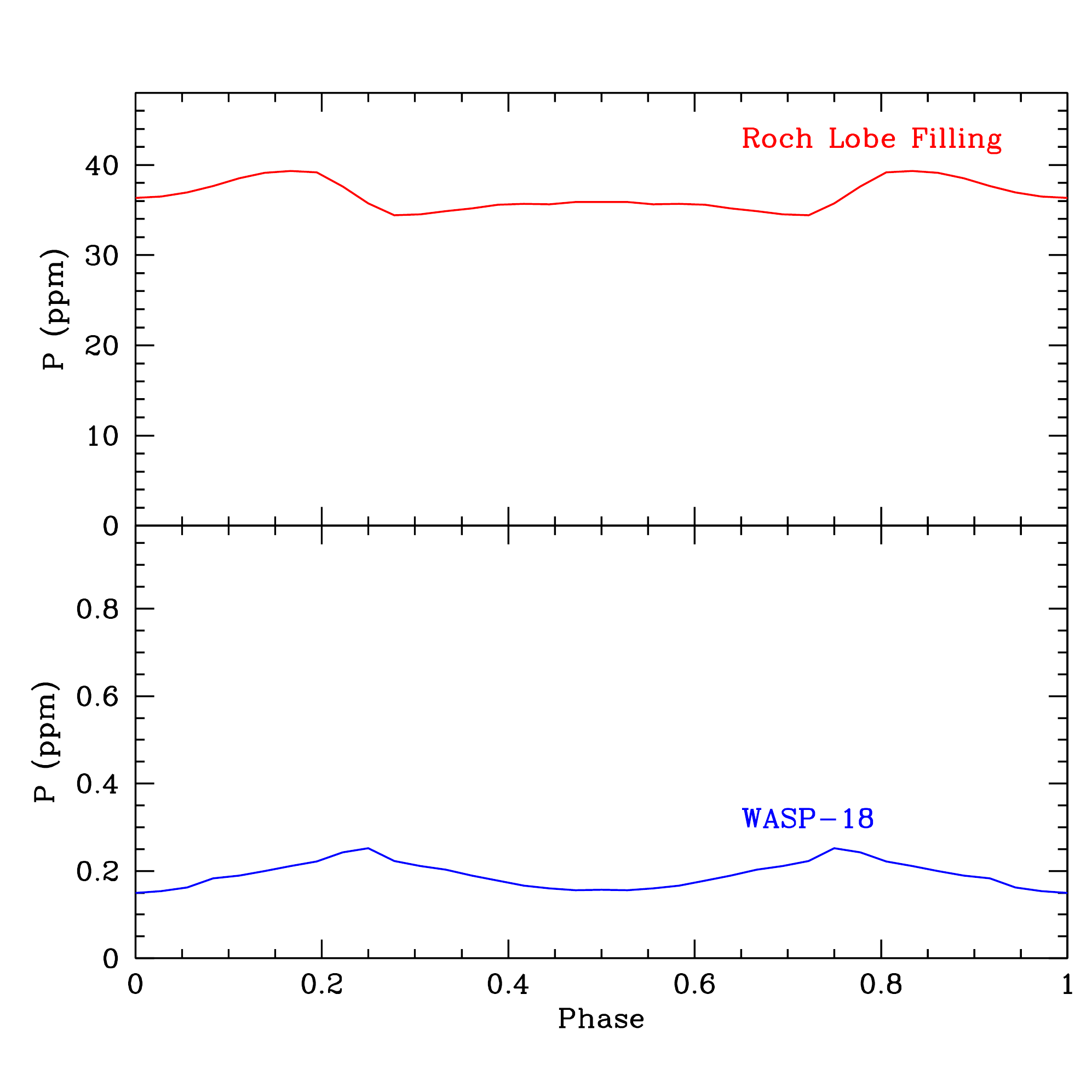}
     \caption{Modelled polarization phase curves at 440 nm due to tidal effects for the two cases shown in Figure \ref{tidal_pol}. The variation in polarized light from this effect for WASP-18 is below HIPPI's sensitivity.}
     \label{tidal_phase}
\end{figure}     

For WASP-18 the mass ratio $q = M_P/M_S $ is 0.0078. We consider first the extreme case of a star large enough to fill its Roche lobe. This requires a star more than twice as large as WASP-18. In a system with such an extreme mass ratio the star has an oblate shape arising from its phase locked rotation, and the tidal distortion mostly shows up in the region close to the L$_1$ point where there is a low gravity region, forming a cold spot on the star (due to gravity darkening) immediately below the planet. This can be seen in Figure \ref{tidal_pol}. The cold spot moves across the star as the phase angle changes and
produces polarization similar to that for the transiting case with the polarization being highest when the cold spot is near the limb which occurs at 60-70 degree phase angles.  For the planet, because the polarized light is reflected light, the oblateness of the planet does not have as great an effect on the path geometry and hence we can expect this is an even smaller, negligible effect.

The resulting phase variations are shown in Figure \ref{tidal_phase}.
The largest polarization effect arises from the oblateness of the star. This produces a phase independent polarization
of 36 ppm at 440 nm for the Roche lobe filling case which is far more extreme than the actual estimated oblateness of WASP 18. The changing position of the cold spot with phase produces a phase
dependent polarization of a little under 4 ppm for the Roche lobe filling case.

In the actual case of WASP-18 the star is much smaller than the Roche lobe, and it is also apparent from the measured $v\sin{i}$ that it is not synchronously rotating. In our model we consider a star rotating 5 times slower than the orbital period. For such a star there is little rotational oblateness, but low-gravity cold spots can be seen on the sides of the star towards and away from the planet. However, the contrast in gravity between the cold spots and the rest of the star is now very small as can be seen in the top right panel of Figure \ref{tidal_pol}. The resulting polarization phase curve (Figure \ref{tidal_phase}) for such a star has an amplitude of only about 0.1 ppm, far below our measurement errors.

Thus it seems that even for this relatively extreme case (as compared with other hot Jupiters) the tidal effects do not
significantly contribute to observed polarization, and this effect can be safely ignored in future studies of exoplanet
polarization at least until the sensitivity of polarimeters is improved.

\subsection{Forward Polarimetric Model}
We have modelled the reflected light polarization of WASP-18b using the VSTAR (Versatile Software for Transfer of Atmospheric Radiation) code \citep{bailey12}  which has been updated to model polarization as described by \citet{bailey18}. The atmospheric model we used for WASP-18b has been used to fit the emission spectrum of the planet determined from secondary eclipse observations with HST, Spitzer and ground-based telescopes as described by \citet{kedziora-chudczer2018}. The model uses a temperature profile with an inversion above ~0.1 bar, and assumes solar metallicity and equilibrium chemistry. Absorption due to molecular and atomic lines of 16 species (H$_2$O, CO, CH$_4$, CO$_2$, C$_2$H$_2$, HCN, TiO, VO, Na, K, Rb, Cs, CaH, CrH, MgH and FeH) is included. Additional opacity sources are Rayleigh scattering due to H, He and H$_2$, H$_2$-H$_2$ and H$_2$-He collision-induced absorption, and bound-free and free-free absorption from H, H$^-$ and H$_2^-$. 


To calculate the reflected light polarization from this model we use the VLIDORT vector radiative transfer code \citep{spurr06} to solve the polarized radiative transfer equation at a grid of pixels spaced across the visible disk of the planet. We then integrate over these pixels to obtain the disk-integrated intensity and polarization from the planet. The methods and a number of verification tests are described in detail by \citet{bailey18}.

\begin{figure*}
    \includegraphics[angle=0,width=\columnwidth,trim={0 5cm 0 5cm},clip]{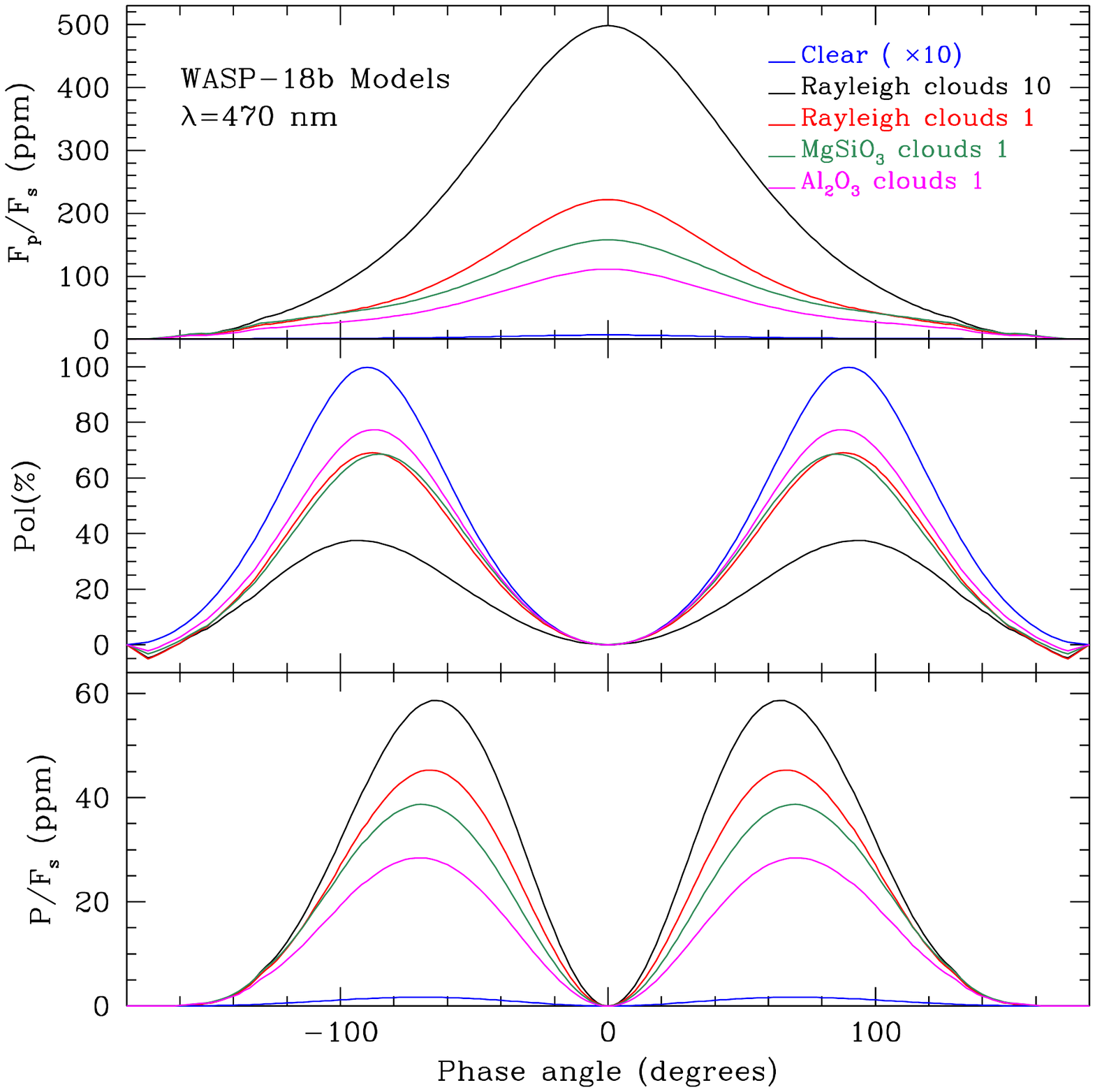}
    \caption{Predictions of the anticipated polarized light curves based on models in \citet{kedziora-chudczer2018} based on new Gemini observations of the system. These models are for a clear atmosphere and various cloud layers. The upper panel is the light reflected from the planet as a fraction of that from the star. The middle panel is the percentage polarization of the light reflected from the planet. The lower panel is the polarization as a fraction of the light from the star and corresponds to the quantity we actually observe. The cloud models are for Rayleigh scattering clouds with optical depths 10 and 1, and clouds with optical depth 1 composed of 0.05 $\mu$m radius enstatite (MgSiO$_3$) and corundum (Al$_2$O$_3$) particles. The clear model has been scaled up by a factor of 10 to be visible at all in the top and bottom panels.}
    \label{vstar_curve}
\end{figure*}

The results for WASP-18b for a wavelength of 470 nm are shown in Figure \ref{vstar_curve} and summarized in Table \ref{lucynafits}. For a model with a clear (cloud free) atmosphere, the reflected light from the planet is extremely weak and the corresponding polarization amplitude is only 0.17 ppm. This is because, although polarized reflected light can be produced by Rayleigh scattering from H$_2$, the absorption opacities (due, for example, to TiO and H$^-$ free-free) are so dominant at this wavelength that very little scattered light is seen.  Some of these absorbers, such as H$^-$ free-free are anticipated to be a substantial part of the atmosphere \citep{Arcangeli2018}. Although the scattered light shows almost 100\% peak polarization (since predominantly single scattering is seen) it remains a very small fraction of the light from the star. The extreme low level of reflected light is different to the case of the cooler planet HD 189733b considered by \citet{bailey18} where a polarization of ~7 ppm could be produced in the clear case.

The Rayleigh scattering cross section of H$_2$ depends strongly on the wavelength with the polarization from H$_2$ (which is the primary source of polarization in the “clear” atmosphere case and the dominant gas in the other cloud cases) increasing towards the UV, however this is a small effect beyond the sensitivity of our measurements.  To quantify this effect, we compared the five atmospheric models---clear atmosphere (H$_2$ dominated), corundum, enstatite, and two ``pure-Rayleigh'' scattering clouds---using our bandpass model in three different spectral bands.  Those bands are: clear pass, which was used for the observations presented here; 500SP, which is HIPPI's primary blue filter used for the HD 189733b observations \citep{Bott2016}; and Johnson U, as an example of a U-band filter.  In the clear atmosphere case the measurements of polarization are estimated to be 0.18, 0.22, and 0.25 ppm for the three bandpasses respectively.  In the other cloud models the values vary by a few parts-per-million at most, with efficiencies of 84\%, 80.2\% and 39.5\% in Clear, 500SP and Johnson U respectively.  In cases where there is an increase in polarization it is not significant enough to overcome the loss in stellar flux at bluer wavelengths.  Multi-band observations could help to distinguish the stellar effects from those of the planet particularly if the star appears to be active (\citep{cotton18} note that WASP 18 is not), however, for the measurements presented here, our bandpass model suggests those variations are beyond our sensitivity.

\begin{table}
\center
\caption{A comparison of the albedos and maximum polarization signal for the models plotted in Figure \ref{vstar_curve}.}
\begin{tabular}{lll}
\hline
Model & Geometric Albedo & Max polarization (ppm) \\
\hline
Clear & 0.00090 & 0.17 \\
Rayleigh clouds ($\tau$ = 10) & 0.673 & 58.7 \\
Rayleigh clouds ($\tau$ = 1) & 0.301 & 45.3 \\
Enstatite clouds ($\tau$ = 1) & 0.214 & 38.7 \\
Corundum clouds ($\tau$ = 1) & 0.150 & 28.4 \\
\hline
\end{tabular}
\label{lucynafits}
\end{table}

Substantial reflected light and polarization can be achieved by including a cloud layer high in the atmosphere. As discussed by \citet{bailey18} the highest polarization is achieved with an optically thick Rayleigh scattering atmosphere. Our model with an optical depth $\tau = 10$ of pure Rayleigh particles produces a polarization amplitude of 58.7 ppm.  These can be thought of as idealized small (compared with the wavelength) non-absorbing particles with a single-scattering albedo of one and a Rayleigh phase function. A similar model with $\tau = 1$ gives 45.3 ppm. Models using 0.05 micron radius particles of enstatite (MgSiO$_3$) and corundum (Al2O$_3$) give slightly lower polarizations (here the particles are ``Rayleigh like'' but include significant absorption unlike our "pure Rayleigh particle" clouds). Both these species---enstatite and corundum---have condensation temperatures too low for the atmospheric temperature profile we are using. However the possibility that clouds form on the colder night-side and are transported to the day-side is considered by \citet{kedziora-chudczer2018}.

The most realistic scenarios here do not exceed the 40 ppm signal ruled out by the polarimetric observations in this paper.  In fact, most cloudy scenarios except Rayleigh scattering particles with a greater optical depth, and certainly a dark clear atmosphere, may produce signals below this threshold.

The model results shown in Figure \ref{vstar_curve} and Table \ref{lucynafits} are monochromatic calculations at 470 nm, while our actual band is quite broad. If polarization varies strongly with wavelength these calculations may not accurately represent what we should observe in our broad band. To test this effect we carried out a full set of wavelength dependent polarization models over the range 300 nm to 1000 nm for one phase angle, 67 degrees, near where the observed polarization peaks, and used our bandpass model to determine the integrated polarization we should observe over the band for each of the models shown in Figure \ref{vstar_curve} and Table \ref{lucynafits}. The polarizations at this phase are 0.18 ppm for the clear model, 59.0 ppm and 41.9 ppm for the two Rayleigh models, and 36.3 ppm and 26.3 ppm for the enstatite and corundum models respectively. These are quite close to the values at this phase angle seen in Figure \ref{vstar_curve}, indicating that no major error is introduced by using a monochromatic model to represent our observations.


\section{Discussion} \label{sec:discuss}





Typically hot Jupiters are selected for polarimetric observation based on the large size of their scattering disk and their proximity to their host star (and hence the amount of light to be scattered).  WASP-18b is an extreme example of this as it orbits especially close to an F-type star.  We are able to produce a constraint on the polarimetric signal from the planet at no greater than 40 ppm from observation, which rules out atmospheres with optically thick ($\tau = $ 10) Rayleigh scattering clouds, and detect an offset likely primarily due to the interstellar medium.  We were unable to detect a polarized light signal modulated by the planet WASP-18b system in spite of its advantages as a candidate.  Possible explanations for this non-detection include a detectable signal just outside of our sensitivity given our observing time, or the planet having a low albedo.

Extended observations or observations with a larger light collecting area can drive down our noise so that we are limited by HIPPI's precision.  For HIPPI or any other high precision polarimeters reaching this sensitivity, obtaining errors in the parts-per-million range could open up the possibilities of determining the cloud species, confirming a dichotomy of albedos on hot Jupiters, and exploring the other phenomena such as the transit.  For stars that are more active than WASP 18, observing the system in multiple wavelength bands may help disentangle the stellar contributions from effects like line-blanketing from stellar activity from the planetary signal.

The seminal paper on which estimates of the polarized light signals of hot Jupiters are typically based, \citet{Seager2000}, concentrated on planets with temperatures half of what has been measured for WASP-18b \citep{nymeyer11} and modelled atmospheres with clouds or strong Rayleigh scattering. Some hot Jupiters may have dark atmospheres without reflective clouds.  WASP-18b is likely one of them considering its high temperatures and the absorbing species likely present in its atmosphere \citep{burrows08, fortney08, zahnle09, knutson10}. For a super hot giant like WASP-18b the atmosphere may be clear on the day-side unless mixing is able to transport clouds from the nightside \citep{kedziora-chudczer2018}. \citet{nymeyer11} found from secondary eclipse observations that the planet is far brighter in infrared than the predicted equilibrium temperature requiring it to have a near-zero albedo and ineffiencent energy transport between the night- and day-sides. 

Therefore, a non-detection of polarized light from this exoplanet should not be regarded as a suggestion that such signals do not exist at the levels predicted by \citet{Seager2000}.  Rather more time for polarimeters capable of detecting at the parts-per-million level, or larger collecting areas, would expedite and enable polarized light detection and characterization of some hot Jupiters, while some other hot Jupiters might lack the thick polarizing or high altitude clouds that would produce a larger signal.  Even in cases of cooler planets which might more readily form condensates, the planets may sometimes be dark enough to only produce signals at a few parts-per-million. 

In the case of HD 189733b the limits placed on the polarized signal from \citet{Bott2016} were a good match to the observations of secondary eclipse in non-polarized photometry in blue as the planet is a deep dark blue color.  In the case of WASP 18b, the color was not known at the time the observations began and relatively less was constrained about its atmosphere than in the HD 189733b case.  The low polarimetric signals from clouds as shown in Figure \ref{vstar_curve} could potentially be observed on a larger telescope, but they are not anticipated to form on the dayside of the planet and would require mixing \citep{kedziora-chudczer2018} to transport them there, although secondary eclipse measurements suggest the planet is quite dark on the dayside \citep{nymeyer11}.  Other hot Jupiters in longer orbits or around cooler stars may be better objects for polarimetric study despite a decrease in incident flux particularly in blue, if they are able to produce reflective clouds. 

The apparent darkness of the planet's dayside from both these polarimetric and previous photometric studies, and \citet{nymeyer11}'s finding that there likely is little, if any, energy redistribution from the dayside to the nightside of the planet suggest mixing is not bringing substantial clouds to the dayside. 
The clouds modelled here and in \citet{kedziora-chudczer2018} and indeed in much of the literature are based largely on refractive indices sources  \citep[e.g][]{scott96, koike95, johnson74, ordal88} for solid phases appropriate to some but not all clouds on hot Jupiters; we caution the community in this regard. Whilst we acknowledge that the high temperatures required to obtain optical data of these materials at liquid phases in a lab is difficult and thus is difficult to find, these solid phase clouds will have optical properties different from what would realistically be found on the dayside of these hottest hot Jupiters.  The models here that include clouds are only appropriate if substantial circulation bringing clouds from the nightside to the dayside is present on WASP-18b and the nightside reaches temperatures significantly lower than on the dayside (e.g. less than about 2300 K at 1 bar for solid ice clouds of Al$_2$O$_3$) and even so, liquid clouds are likely to also be present.


\section{Conclusions} \label{sec:conclusions}

We have described new measurements in linearly polarized light of the hot Jupiter exoplanet hosting system WASP-18, made with the HIPPI polarimeter.  The system provides an unusual test case in the form of a high mass hot Jupiter (M $\approx 10$ M$_{J}$) orbiting very closely to its star ($p < 1$ day).  

Our best fits to a Rayleigh Lambertian phase curve are not changed substantially by the inclusion of
transit observations.  The best fit to the modulation in the observations is for  WASP-18b to orbit with
a position angle of $200.3 \pm 20.7$ degrees and an inclination of  $79.2 \pm 10.9$ degrees, which is in
agreement with literature although low values of inclination within that range can be omitted since it
is known that the planet transits.  With fitted offsets of Z$_q$ and Z$_u$ at $-75.2\pm6.0$ ppm and
$185.0\pm5.5$ ppm, the best fit to the signal itself is for a polarization of $16.2\pm10.0$ ppm from the
planet which is in good agreement with predictions for polarized light signals from hot Jupiters.  

The offsets from zero are to the order of those expected from the interstellar medium and are not expected to come from contributions of the star or circumstellar debris.  

We find the signal is statistically not differentiable from a best fit to noise: it is possible to produce a fit to noise for signals of this amplitude with this level of sampling.  We can, however, with 99\% confidence, rule out a detection at the 40 ppm level. Forward models for the expected polarization signal for the planet exceed our 40 ppm limit only for the most extraordinary cases of pure Rayleigh scattering clouds. Models with more realistic cloud particles can produce polarization amplitudes with 26-36 ppm amplitudes which are still well above our best fit polarization amplitude. The most likely reason that we are not seeing a planetary polarization signal is that the atmosphere of WASP-18b is too hot for clouds to form or move to the day-side. A clear atmosphere results in a very low geometric albedo and a polarization well below our detection limits.

Other noise sources and polarization mechanisms were explored.  Tides are inflicted on the star and planet in this
tight orbit, however we find they are unlikely to produce a detectable signal.  We also assess the transit signal in
polarized light and find that we do not detect the effects predicted by \citet{wiktorowicz14}.

This paper serves as a template for the thorough treatment of polarized light signals from exoplanets.  The observational approach is promising but relatively novel and so requires the thoughtful treatment of noise sources and combined polarimetric effects.  Further observations on an 8-meter class telescope with a HIPPI analogue instrument should produce substantial data points to provide confidence in either a detection under 40 ppm of polarized light from Rayleigh scattering from the planet, or a firm non-detection.

\acknowledgments

The development of HIPPI was funded by the Australian Research Council through Discovery Projects
grant DP140100121 and by the UNSW Faculty of Science through its Faculty Research Grants program. The
authors thank the Director and staff of the Australian Astronomical Observatory for their advice and 
support with interfacing HIPPI to the AAT and during the observing runs on the telescope.  The analyses of these observations were reliant upon data from the SIMBAD database and Gaia archive.  KMB and VSM acknowledge support from the NASA Astrobiology Institutes's Virtual Planetary Laboratory, funded under cooperative agreement number NAA13AA93A.

%

\vspace{5mm}
\facilities{AAT, UNSW: HIPPI}


\software{  
          VSTAR \citep{bailey12,bailey18}, 
          VLIDORT \citep{spurr06}, 
          SYNSPEC \citep{hubeny85}
          }


\begin{thebibliography}{99}
\bibitem[\protect\citeauthoryear{Arcangeli et al.}{2018}]{Arcangeli2018} Arcangeli, J., Desert, J.-M., Line, M.R., Bean, J.L., Parmentier, V., Stevenson, K.B., Kreidberg, L., Fortney, J.J., Mansfield, M., \& Showman, A.P., ApJ Letters, 855, 2, 272
\bibitem[\protect\citeauthoryear{Bailey}{2007}]{bailey07} Bailey J., 2007, Astrobiology, 7, 2, 320-332
\bibitem[\protect\citeauthoryear{Bailey et al.}{2010}]{bailey10} Bailey J., Lucas P.W., \&
Hough J.H., 2010, MNRAS, 405, 2570
\bibitem[\protect\citeauthoryear{Bailey \& Kedziora-Chudczer}{2012}]{bailey12} Bailey J. \& Kedziora-Chudczer, L., 2010, MNRAS, 419, 3, pp. 1913-1929
\bibitem[\protect\citeauthoryear{Bailey et al.}{2015}]{bailey15} Bailey J., Kedziora-Chudczer L., Cotton D.V., Bott K., Hough J.H., \& Lucas P.W., 2015, MNRAS, 449, 3064
\bibitem[\protect\citeauthoryear{Bailey et al.}{2018}]{bailey18} Bailey J., Kedziora-Chudczer L., \& Bott K., 2018, MNRAS, 480, 1613-1625
\bibitem[\protect\citeauthoryear{Berdyugina et al.}{2011}]{berdyugina11} Berdyugina S.V., Berdyugin A.V., Fluri D.M., Piirola V., 2011, ApJL, 728, L6
\bibitem[\protect\citeauthoryear{Bott et al.}{2016}]{Bott2016} Bott, K., Bailey, J., Kedziora-Chudczer, L., Cotton, D.V., Lucas, P.W., Marshall, J., \& Hough, J.H, 2016, MNRAS, 459, 1, pL109-L113 
\bibitem[\protect\citeauthoryear{Bourrier et al.}{2013}]{Bourrier2013} Bourrier, V., Lecavelier des Etangs, A., Dupuy, H., Ehrenreich, D., Vidal-Madjar, A., Hebrard, G., Ballester, G.E.., Desert, J.M., Ferlet, R., Sing, D.K., \& Wheatley, P.J., 2013, A\& A, 551, A63
\bibitem[\protect\citeauthoryear{Brooks et al.}{1994}]{brooks94} Brooks, A., Clarke, D., \& McGale, P.A., 1994, Vistas in Astronomy, 38, 4, 377-399
\bibitem[\protect\citeauthoryear{Brown et al.}{2011}]{Brown2011} Brown, D.J.A., Collier Cameron, A., Hall, C., Hebb, L., \& Smalley, B., 2011, MNRAS, 415, 1, pp. 605--618
\bibitem[\protect\citeauthoryear{Burrows et al.}{2008}]{burrows08} Burrows, A., Ibgui, L., \& Hubby, I., 2008, ApJ 682, 1277
\bibitem[\protect\citeauthoryear{Carciofi \& Magalhaes}{2005}]{carciofi05} Carciofi, A.C. \& Magalhaes, A.M., 2005,ApJ, 635, 570
\bibitem[\protect\citeauthoryear{Castelli \& Kurucz}{2004}]{castelli04} Castelli, F. \& Kurucz, R.L., 2004, arXiv:astroph/0405087
\bibitem[\protect\citeauthoryear{Chandrasekhar}{1946}]{chandrasekhar46} Chandrasekhar, S., 1946, ApJ, 103, 351
\bibitem[\protect\citeauthoryear{Cotton et al.}{2016}]{cotton16} Cotton D.V., Bailey J., Kedziora-Chudczer L., Bott K., Lucas P.W., Hough J.H., \& Marshall J.P., 2016, MNRAS, 455, 1607
\bibitem[\protect\citeauthoryear{Cotton et al.}{2017a}]{cotton17a} Cotton, D.V., Marshall, J.P., Bailey, J., Kedziora-Chudczer, L., Bott, K., Marsden, S.C., \& Carter, B.D., 2017(a), MNRAS, 467, 1, p. 873-897
\bibitem[\protect\citeauthoryear{Cotton et al.}{2017b}]{cotton17b} Cotton, D.V., Bailey, J., Howarth, I.D., Bott, K., Kedziora-Chudczer, L., Lucas, P.W., \& Hough, J.H., 2017(b), Nature Ast, 1, 690
\bibitem[\protect\citeauthoryear{Cotton et al.}{2018}]{cotton18} Cotton, D.V., Evensberget, D., Marsden, S.C., Bailey, J., Zhao, J., Kedziora-Chudczer, L., Carter, B.D., Bott, K., Vidotto, A.A., Petit, P., Morin, J., \& Jeffers, S.V. 2018, submitted to MNRAS
\bibitem[\protect\citeauthoryear{de Kok \& Stam}{2012}]{dekok12} de Kok, R.J. \& Stam, D.M., 2012, Icarus, 221, 2, pp.517-524
\bibitem[\protect\citeauthoryear{Espinosa Lara \& Rieutord}{2012}]{espinosa12} Espinosa Lara, F. \& Rieutord, M., 2012, A\&A, 547, A32
\bibitem[\protect\citeauthoryear{Fares et al.}{2010}]{fares10} Fares, R., Donati, J.F., Moutou, C., Jardine, M.M., Griemeier, J.M., Zarka, P., Shkolnik, E.L., Bohlender, D., Catala, C., \& Collier Cameron, A., 2010, MNRAS, 406, 409-419
\bibitem[\protect\citeauthoryear{Fauchez et al.}{2017}]{fauchez17} Fauchez, Th., Rossi, L., \& Stam, D.M., 2017, ApJ, 842, 41, 18pp
\bibitem[\protect\citeauthoryear{Fluri \& Berdyugina}{2010}]{fluri10} Fluri, D.M., \& Berdyugina, S.V., 2010, A\&A, 512, A59
\bibitem[\protect\citeauthoryear{Fortney et al.}{2008}]{fortney08} Fortney, J.J., Lodders, K., Marley, M.S., \& Freedman, R.S., 2008, ApJ 678, 1419
\bibitem[\protect\citeauthoryear{The Gaia Collaboration et al.}{2016}]{gaia16} The Gaia Collaboration et al., 2016, A\&A, 595, A1, 36 pp.
\bibitem[\protect\citeauthoryear{Gaudi et al.}{2017}]{gaudi17} Gaudi, S.B., Stassun, K.G., Collins, K.A., Beatty, T.G., Zhou, G., \& 55 other coauthors, 2017, Nature, 546, 514
\bibitem[\protect\citeauthoryear{Haswell et al.}{2012}]{Haswell2012} Haswell, C.A., Fossati, L., Ayres, T., France, K., Froning, C.S., Holmes, S., Kolb, U.C., Busuttil, R., Street, R.A., \& Hebb, L., 2012, ApJ, 760, 1, 79
\bibitem[\protect\citeauthoryear{Hellier et al.}{2009}]{Hellier2009} Hellier, C., Anderson, D.R., Cameron, A. C., Gillon, M, Hebb, L., Maxted, P.F.L., Queloz, D., Smalley, B., Triaud, A.H.M.J., \& West, R., et al, 2009, Nature, 460, 1098--1100
\bibitem[\protect\citeauthoryear{Hough \& Lucas}{2003}]{hough03} Hough, J.H. \& Lucas, P.W., 2003, in Fridlund, M., Henning, T., eds. Towards other Earths: DARWIN/TPF and the Search for Extrasolar Terrestrial Planets, ESA SP-539, ESA Publications Division, Noordwijk, p. 11 
\bibitem[\protect\citeauthoryear{Hough et al.}{2006}]{hough06} Hough J.H., Lucas P.W., Bailey
J.A., Tamura M., Hirst E., Harrison D., \& Bartholomew-Biggs M, 2006, PASP, 118, 1302
\bibitem[\protect\citeauthoryear{Hubeny et al.}{1985}]{hubeny85} Hubeny, I., Stefl, S., \& Harmanec, P., 1985, Bull. Astron. Inst. Czechosl., 36, 214
\bibitem[\protect\citeauthoryear{Johnson \& Christy et al.}{1974}]{johnson74} Johnson, P.B., \& Christy, R.W., 1974, Phys. Rev. B, 9, 5056
\bibitem[\protect\citeauthoryear{Karalidi et al.}{2013}]{karalidi13} Karalidi, T., Stam, D.M., \& Guirado, D., 2013,  A\&A, 555, A127
\bibitem[\protect\citeauthoryear{Kedziora-Chudczer et al.}{2018}]{kedziora-chudczer2018} Kedziora-Chudczer, L., Zhou, G., Bailey, J., Bayliss, D.D.R., Tinney, C.G., \& Osip, D., 2018, MNRAS, submitted
\bibitem[\protect\citeauthoryear{Kemp \& Barbour}{1981}]{kemp81} Kemp J. C. \& Barbour M.S., 1981,  PASP, 93, 521-525.
\bibitem[\protect\citeauthoryear{Kemp et al.}{1983}]{kemp83} Kemp, J.C., Henson, G.D., Barbour, M.S., Kraus, D.J., \& Collins, G.W., 1983, ApJL, 273, L85
\bibitem[\protect\citeauthoryear{Kemp et al.}{1987}]{kemp87} Kemp, J.C., Henson, G.D., Steiner, C.T., \& Powell, E.R., 1987, Nature, 326, pp 270--273
\bibitem[\protect\citeauthoryear{Knutson et al.}{2010}]{knutson10} Knutson, H.A., Howard, A.W., \& Isaacson, H., 2010, ApJ 720, 1569
\bibitem[\protect\citeauthoryear{Kolokolova \& Kimura}{2010}]{kolokolova10} Kolokolova, L. \& Kimura, H., 2010, A\&A, 513, A40
\bibitem[\protect\citeauthoryear{Koike et al.}{1995}]{koike95} Koikem C., Kaito, C., Yamamoto, T., Shibai, H., Kimura, S., \& Suto, H., 1995, Icarus, 114, 203
\bibitem[\protect\citeauthoryear{Kostogryz et al.}{2011}]{kostogryz11} Kostogryz, N.M., Yakobchuk,
T.M., Morzhenko, O.V., \& Vid'machenko, A.P., 2011, MNRAS, 415, 695 
\bibitem[\protect\citeauthoryear{Kostogryz et al.}{2015}]{kostogryz15} Kostogryz N.M., Yakobchuk T.M., \& Berdyugina S.V., 2015, ApJ, 806, 1, 97
\bibitem[\protect\citeauthoryear{Linsky et al.}{2010}]{Linsky2010} Linksky, J.L., Yang, H., France, K., Froning, C.S., Green, J., Stocke, J.T., \& Osterman, S.N., 2010, ApJ, 717, 2, p 1291
\bibitem[\protect\citeauthoryear{Lucas et al.}{2009}]{lucas09} Lucas P.W., Hough J.H., Bailey J.A., Tamura M., Hirst E., Harrison D., 2009, MNRAS, 403, 4, pp 1949-1968
\bibitem[\protect\citeauthoryear{Leroy}{1999}]{leroy99} Leroy, J.L., 1999, A\&A, 346, pp. 955-60
\bibitem[\protect\citeauthoryear{Marshall et al.}{2016}]{marshall16} Marshall, J., Cotton, D.V., Bott, K., Ertel, S., Kennedy, G.M., Wyatt, M.C., del Burgo, C., Absil, O., Bailey, J., \& Kedziora-Chudczer, L., 2016, ApJ, 825, 2, id.124
\bibitem[\protect\citeauthoryear{Maxted et al.}{2013}]{Maxted2013} Maxted, P.F.L., Anderson, D.R., Doyle, A.P., Gillon, M,. Harrington, J., Iro, N., Jehin, E., Lafreni{\`e}re, D., Smalley, B., \& Southworth, J., 2013, MNRAS, 428, pp 2645--2660
\bibitem[\protect\citeauthoryear{Miller et al.}{2012}]{miller12} Miller, B.P., Gallo, E., Wright, J.T., \& Dupree, A.K., 2012, ApJ, 754, 2, id. 137
\bibitem[\protect\citeauthoryear{Nymeyer et al.}{2011}]{nymeyer11} Nymeyer, S., Harrington, J., Hardy, R.A., Stevenson, K.B., Campo, C.J., Madhusudhan, N., Collier-Cameron, A., Loredo, Th., Blecic, J., \& Bowman, W.C., 2011, ApJ, 742, 1, id. 35
\bibitem[\protect\citeauthoryear{Ordal et al.}{1988}]{ordal88} Ordal, M.A., Bell, R.J., Alexander, R.W., Newquist, L.A., \& Querry, M.R., 1988, Applied Opt., 27, 1203
\bibitem[\protect\citeauthoryear{Perryman et al.}{1997}]{perryman97} Perryman, M.A.C., Lindegren, L., Kovalevsky, J., Hoeg, E., Bastian, U., Bernacca, P.L., Creze, M., Donati, F., Grenon, M., Grewing, M., van Leeuwen, F., van der Marel, H., Mignard, F., Murray, C.A., Le Poole, R.S., Schrijver, H., Turon, C., Arenou, F., Froeschle, M., \& Petersen, C.S., 1997, A\&A, 323, p.L49-L52
\bibitem[\protect\citeauthoryear{Pillitteri et al.}{2014}]{pillitteri14} Pillitteri, I., Wolk, S.J., \& Sciortino, S., 2014, A\&A, Vol.567, A128
\bibitem[\protect\citeauthoryear{Press et al.}{1992}]{press92} Press W.H., Teukolsky S.A., Vetterling W.T., \& Flannery B.P., 1992, {\it Numerical Recipes in FORTRAN: The Art of Scientific Computing}, Cambridge University Press
\bibitem[\protect\citeauthoryear{Roberge et al.}{2012}]{roberge12} Roberge, A., Chen, C., Millan-Gabet, R., Weinberger, A.J., Hinz, Ph.M., Stapelfeldt, K.R., Absil, O., Kuchner, M., \& Bryden, G., 2012, PASP, 124, 918
\bibitem[\protect\citeauthoryear{Salz et al.}{2015}]{Salz2015} Salz, M., Schneider, P.C., Czesla, S., \& Schmitt, J.H.M.M., 2015, A\&A, 576, A42
\bibitem[\protect\citeauthoryear{Scott \& Duley}{1996}]{scott96} Scott, A, \& Duley, W.W., 1996, ApJ Supplement, 105, 401
\bibitem[\protect\citeauthoryear{Seager et al.}{2000}]{Seager2000} Seager S., Whitney B.A., \& Sasselov D.D., 2000, ApJ, 540, 504
\bibitem[\protect\citeauthoryear{Simmons \& Stewart}{1985}]{simmons85} Simmons, J.F.L. \& Stewart, B.G., 1985, A\&A, 142, 100
\bibitem[\protect\citeauthoryear{Skrutskie et al.}{2006}]{skrutskie06} Skrutskie,M.F., Cutri, R.M., Stiening, R., Weinberg, M.D., Schneider, S., Carpenter, J.M., \& 25 other authors, 2006, ApJ, 131, 2, pp 1163-1183
\bibitem[\protect\citeauthoryear{Southworth et al.}{2009}]{Southworth2009} Southworth J., Hinse, T.C., Dominik, M., Glitrup, M.,  J{\o}rgensen, U.G., Liebig, C., Mathiasen, M., Anderson, D.R., Bozza, V., Browne, P., et al, 2009, ApJ, 707, 1, pp. 167
\bibitem[\protect\citeauthoryear{Spurr}{2006}]{spurr06} Spurr, R., 2006, \jqsrt, 102, 316.
\bibitem[\protect\citeauthoryear{Stam et al.}{2004}]{stam04} Stam, D.M., Hovenier, J.W., \& Waters, L.B.F.M., 2004, A\&A, 428, 663–672
\bibitem[\protect\citeauthoryear{Torres et al.}{2008}]{Torres2008} Torres G., Winn J.N., \& Holman M.J., ApJ, 677, 1324-1342
\bibitem[\protect\citeauthoryear{Torres et al.}{2012}]{torres12} Torres, G., Fischer, D.A., Sozzetti, A., Buchave, L.A., Winn, J.N., Holman, M.J., \& Carter, J.A., 2012, ApJ, 757, 161
\bibitem[\protect\citeauthoryear{Von Zeipel}{1924}]{vonzeipel24} Von Zeipel, H., 1924, MNRAS, 84, p.665-683
\bibitem[\protect\citeauthoryear{Walker}{2008}]{walker08} Walker, G.A.H., Croll, B., Matthews, J.M., Kuschnig, R., Huber, D., Weiss, W.W., Shkolnik, E., Rucinski, S.M., Guenther, D.B., Moat, A.F.J., \& Sasselov, D., 2009, A\&A, 482, 2, 691-697
\bibitem[\protect\citeauthoryear{Wiktorowicz}{2009}]{wiktorowicz09} Wiktorowicz S.J., 2009, ApJ, 696, 1116
\bibitem[\protect\citeauthoryear{Wiktorowicz \& Matthews}{2008}]{wiktorowicz08} Wiktorowicz S.J.   \& Matthews K., 2008, PASP, 120, 1282
\bibitem[\protect\citeauthoryear{Wiktorowicz \& Laughlin}{2014}]{wiktorowicz14} Wiktorowicz S. \& Laughlin, G.P., 2014, ApJ, 795, 1, id 12
\bibitem[\protect\citeauthoryear{Wiktorowicz \& Nofi}{2015}]{wiktorowicz15a} Wiktorowicz S. \&  Nofi, L.A., 2015, ApJ, 800, L1
\bibitem[\protect\citeauthoryear{Wiktorowicz et al.}{2015}]{wiktorowicz15b} Wiktorowicz S., Nofi L.A., Daniel J.-T., Kopparla P., Laughlin G.P., Hermis N., Yung Y.L., \& Swain M.R., 2015, ApJ, 813, 48
\bibitem[\protect\citeauthoryear{Wilkins et al.}{2017}]{Wilkins2017} Wilkins, A., Delrex, L., Barker, A., Deming, D., Hamilton, D., Gillon, M., \& Jehin, E., 2017, ApJ Letters, 836, 2, L24
\bibitem[\protect\citeauthoryear{Wilson \& Devinney}{1974}]{wilson74} Wilson, R.E \& Devinney, E.J., 1974, ApJ, 166, 605
\bibitem[\protect\citeauthoryear{Wilson}{1979}]{wilson79} Wilson, R.E., 1979, ApJ 234, 1054
\bibitem[\protect\citeauthoryear{Wright et al.}{2010}]{wright10} Wright, E.L., Eisenhardt, P.R.M., Mainzer, A.K., Ressler, M.E., Cutri, R.M., Jarrett, Th., Kirkpatrick, J.D., Padgett, D. \& 30 other authors, 2010, ApJ, 140, 6, pp 1868-1881
\bibitem[\protect\citeauthoryear{Zahnle et al.}{2009}]{zahnle09} Zahnle, K., Marley, M.S., Freedman, R.S., Lodders, K., \& Fortney, J.J., 2009, ApJ 701, L20

\end{thebibliography}
\end{document}